\documentclass[preprint]{elsarticle}
\usepackage{fullpage}
\usepackage{listings}

\usepackage{diagbox}
\usepackage{amsmath}
\usepackage[breaklinks=true]{hyperref}
\usepackage{booktabs}
\usepackage{xcolor}
\usepackage{url}
\usepackage[T1]{fontenc}

\AtBeginDocument{%
  \providecommand\BibTeX{{%
    \normalfont B\kern-0.5em{\scshape i\kern-0.25em b}\kern-0.8em\TeX}}}

\begin{document}

\lstset{
  basicstyle=\ttfamily,
  columns=fullflexible,
  keepspaces=true,
}

\title{Automated Translation and Accelerated Solving of Differential Equations on Multiple GPU Platforms}

\author[inst1]{Utkarsh}
\affiliation[inst1]{%
  organization = {Massachusetts Institute of Technology},
  city = {Cambridge},
  state = {Massachusetts},
  country = {USA}
}

\author[inst1]{Valentin Churavy}

\author[inst2]{Yingbo Ma}
\affiliation[inst2]{%
  organization = {JuliaHub},
  city = {Cambridge},
  state = {Massachusetts},
  country = {USA}
}

\author[inst2]{Tim Besard}

\author[inst1]{Prakitr Srisuma}

\author[inst1,inst3]{Tim Gymnich}
\affiliation[inst3]{%
  organization = {Technical University of Munich},
  city = {Munich},
  country = {Germany},
}

\author[inst4]{Adam R.~Gerlach}
\affiliation[inst4]{%
  organization = {United States Air Force Research Laboratory},
  city = {Wright-Patterson AFB},
  state = {OH},
  country = {USA}
}

\author[inst1]{Alan Edelman}

\author[inst1]{George Barbastathis}

\author[inst1]{Richard D. Braatz}

\author[inst1,inst2,inst5]{Christopher Rackauckas}

\affiliation[inst5]{%
  organization = {Pumas-AI},
  city = {Dover},
  state = {Delaware},
  country = {USA}
}

\newcommand{\RDnote}[1]{\textcolor{red}{#1}}


\begin{abstract}

We demonstrate a high-performance vendor-agnostic method for massively parallel solving of ensembles of ordinary differential equations (ODEs) and stochastic differential equations (SDEs) on GPUs. The method is integrated with a widely used differential equation solver library in a high-level language (Julia's DifferentialEquations.jl) and enables GPU acceleration without requiring code changes by the user.
Our approach achieves state-of-the-art performance compared to hand-optimized CUDA-C++ kernels while performing 20--100$\times$ faster than the vectorizing map (\texttt{vmap}) approach implemented in JAX and PyTorch.
Performance evaluation on NVIDIA, AMD, Intel, and Apple GPUs demonstrates performance portability and vendor-agnosticism. We show composability with MPI to enable distributed multi-GPU workflows.
The implemented solvers are fully featured -- supporting event handling, automatic differentiation, and incorporation of datasets via the GPU's texture memory -- allowing scientists to take advantage of GPU acceleration on all major current architectures without changing their model code and without loss of performance. We distribute the software as an open-source library, \texttt{DiffEqGPU.jl}.
\end{abstract}



\maketitle

\section{Introduction}

Solving ensembles of the same differential equation with different choices of parameters and initial conditions is common in many technical computing scenarios such as solving inverse problems \cite{tarantola2005inverse}, performing uncertainty quantification \cite{kuhn2009monte,marino2008methodology,metropolis1949monte}, and calculating global sensitivity analysis \cite{marino2008methodology, iooss2015review}. While such a naturally parallel problem lends itself to being well-suited for acceleration via GPU hardware, the programming requirements have traditionally been a barrier to the adoption of GPU-parallel solvers by scientists and engineers who are less programming savvy. The core difficulty of targeting GPUs with general ODE solver software is that the definition of the ODE is a function given by the user. Thus, high-level ODE solver software has generally consisted of higher-order functions which take as input a function written in a high-level language such as MATLAB \cite{Shampine_1997_MATLAB}, Python (SciPy \cite{virtanen2020scipy}), or Julia (DifferentialEquations.jl \cite{Rackauckas_2017_Julia}) to reduce the barrier to entry for scientists and engineers. In order to target GPUs, previous software such as MPGOS \cite{hegedHus2021program} has required users to rewrite their models in a kernel language such as CUDA C++, which has thus traditionally kept optimized GPU usage out of reach for many scientists. In order to get around this barrier, some software for general GPU-based ODE solving in high-level languages has targeted array-based interfaces such as found in machine learning libraries like PyTorch \cite{li2020scalable} and JAX \cite{jax2018github}. However, we demonstrate in this manuscript that such an approach is orders of magnitude less performant than generating model-specific ODE solver kernels.

In this manuscript, we demonstrate a performant, composable, and vendor-agnostic method for model-specific kernel generation to solve massively parallel ensembles of ordinary differential equations (ODEs) and stochastic differential equations (SDEs) on GPUs. The ODE solvers support both stiff and non-stiff ODEs, allowing a wide range of compatibility with different classes of models. Our software transforms code that targets a widely used differential equation solver library in a high-level language (Julia's DifferentialEquations.jl \cite{Rackauckas_2017_Julia}) and automatically generates optimized GPU kernels without requiring code changes by the end user. We demonstrate an array-based parallelism approach and an automated kernel generation approach, which give a trade-off in extensibility and performance. We demonstrate that the kernel generation achieves state-of-the-art performance by on average outperforming hand-optimized CUDA-C++ kernels provided by MPGOS, and performing $20$--$100\times$ faster than the vectorized map ({vmap}) approach implemented in JAX and PyTorch. We showcase the vendor-agnostic aspect of our approach by benchmarking the results on many major GPU vendors' cards like NVIDIA, AMD, Intel (oneAPI), and Apple Silicon (Metal) and demonstrate the composability with MPI to enable distributed multi-GPU workflows. We show that these solvers are fully featured, supporting event handling, forward and reverse (adjoint) automatic differentiation, and incorporation of datasets via the GPU's texture memory. Together, this software allows scientists to target all major GPU platforms without loss of performance.To summarize, the key contributions of this manuscript are:

\begin{itemize}
    \item A feature-rich open-source library of massively parallel GPU ODE and SDE solvers without trading the high-level interface and performance, allowing composability with the rest of the numerical computing ecosystem.
    \item According to the author's knowledge, the first known implementation to be completely vendor-agnostic provides a roadmap of the required groundwork for numerical software to achieve vendor-agnosticism. 
    \item Increased the performance of GPU parallelized stiff ODE solvers, which is enabled by leveraging and extending automatic differentiation for GPUs by static compilation.
    \item Multiple algorithm choices lead to insights in performance engineering for the problem: We showcase that traditional parallelization of numerical solvers such as LSODA, which are known to be performant for large ODEs, are orders of magnitude slower at solving multiple and small ODEs together on GPUs due to their heavy use of branching behavior. Alternative methods (Rosenbrock methods) are thus demonstrated as more suitable for the use of ensemble GPU parallelism.
\end{itemize}
\section{Related Work}

While researchers have used GPUs to accelerate computations extensively in applications including molecular simulation, biological systems, and physics \cite{zhou2011gpu, fernando2022gpu, zhao2021overcoming, le2013spfp}, these implementations are generally CUDA kernels written for the specific models and thus are not general ODE solver software. In order to simplify the targeting of GPUs with general ODE solver software, previous attempts have generally targeted hardware using array abstraction frameworks such as ArrayFire \cite{malcolm2012arrayfire}, Thrust \cite{bell2012thrust}, VexCL \cite{demidov2012vexcl}, JAX \cite{jax2018github}, and PyTorch \cite{li2020scalable}. These frameworks allow the user to adapt code written on high-level array abstractions and generate a highly optimized code to backends such as  OpenCL \cite{stone2010opencl} and CUDA \cite{nickolls2008scalable}. Boost's odeint \cite{ahnert2011odeint,ahnert2014solving,nagy2022art} allows direct calls to ODE solvers that, without any modification, work with GPU backends such as CUDA and OpenCL.  JAX's Diffrax \cite{kidger2021on} generates solvers for ensembles of ODEs on GPUs via JAX's vectorized map functionality (\texttt{vmap}). PyTorch's torchdiffeq \cite{chen2018neural} allows the defining of ensembles of ODEs directly with GPU-based arrays, although their \texttt{vmap} provided by functorch support with ODEs is still primitive as of April 2023.

Recent results have demonstrated that using array-based abstractions for generating GPU-parallel ODE ensemble solvers greatly lags in performance compared to the state of the art. In particular, MPGOS \cite{hegedHus2021program} demonstrated that ODEINT was 10--100$\times$ slower than purpose-written ODE solver kernels written in CUDA. In order to achieve this performance, MPGOS requires that the user write CUDA C++ kernels for the ODE definitions, which are then compiled into the solver to reduce the kernel call overhead. Similar results were seen with culsoda \cite{zhou2011gpu}, a CUDA translation of the widely used LSODE solver \cite{hindmarsh1983odepack, hindmarsh1995algorithms}, which was similarly limited due to requiring ODE models to be written in CUDA and compiled into the kernels.

\section{Numerical Methods for Differential Equations}

\subsection{Non-stiff Ordinary Differential Equations (ODEs)} \label{section: odes}

ODEs are models given by the evolution equation:
\begin{equation}
    \frac{du}{dt} = f(u,p,t),
\end{equation}
with the initial condition $u(t_0) = u_0$ over the time span $(t_0,t_f)$, where $u$ is the solution, $p$ is the parameter, $t$ is time, $t_0$ is the initial time, and $t_f$ is the final time. In the subsequent sections of this article, the parameter $p$ is omitted from the formulation to avoid confusion as it is not an important part of our numerical methods and algorithms. There exist many different methods for numerically solving ODEs \cite{hairer1993solving, hairer1991solving}, though generally the most performant method is determined by a property known as the stiffness of the ODE, which is related to the pseudo-spectra of the Jacobian \cite{shampine2007stiff, higham1993stiffness}. 

One of the most common classes of solvers for ODE software are explicit Runge-Kutta methods \cite{runge1895numerische, kutta1901beitrag}. These methods are specified by a coefficient tableau $\{A,b,c\}$, with "$s$" stages and order "$k$", where $k\leq s$. It produces the approximation for $u_n = u(t_0 + nh)$ which is the solution at the current time step $t_n$, $h$ is the time step ($t_{n+1}=t_n+h$), where $h$ is the time-step, as
\begin{align}
    k_s &= f\left(u_n + \sum^s_{i=1}a_{s,i}k_i,  t  + c_sh \right) \\
    u_{n+1} &= u_n + h\sum_{i=1}^sb_ik_i
\end{align}
Some  examples of Runge-Kutta methods include dopri5 \cite{dormand1980family} and MATLAB's ODE suite ode45 \cite{Shampine_1997_MATLAB}.

For adaptive step-size control, the Runge-Kutta methods require an extra computation as $\Tilde{u}(t+h) = u(t) + h\sum_{i=1}^s\Tilde{b}_ik_i$, where $\Tilde{b}_i$ are another linear combiners, which approximates the solution by one order less than the original solution. The local error estimate can be written as $E = \lVert \Tilde{u}(t+h) - u(t+h) \rVert$ \cite{hairer1993solving, ascher1998computer}. Adaptivity ensures that error remains below certain tolerance, and these tolerances are absolute (atol) and relative (rtol). Mathematically, the proportion of error against tolerance is
\begin{equation}
    q = \left\| {\frac{E}{\mathrm{atol} + \mathrm{rtol} \cdot \max \{|u(t)|, |u(t+h)|\}}} \right\|.
\end{equation}
The step-size $h$ is accepted for $q<1$; otherwise, $h$ is reduced and a new step is attempted. The new step-size in Runge-Kutta methods is proposed through proportional-integral control (PI-control) via $h_\mathrm{new} = \eta q_{n-1}^{\beta_2}q_n^{\beta_1}h$, where $\beta_1,\beta_2$ are tuned parameters \cite{hairer1993solving}, $q_{n-1}$ is the previous proportion error, and $\eta$ is the safety factor. 

\subsection{Stiff Ordinary Differential Equations}

\subsubsection{Rosenbrock Methods}
There exist various numerical methods for solving stiff ODEs \cite{hairer1991solving,Shampine_1997_MATLAB}. One of the most common algorithms used in various ODE solver packages, e.g., Julia, MATLAB, is known as the {\it Rosenbrock method} \cite{Rosenbrock1963,hairer1991solving,Shampine_1997_MATLAB,Rackauckas_2017_Julia}. The general formulas of an $s$-stage Rosenbrock method is given by
\begin{align}
    &\begin{aligned}
        k_i = &hf\left(u_n + \sum_{j=1}^{i-1} \alpha_{ij}k_j,t_n+\alpha_ih\right) + \beta_ih^2\dfrac{\partial f}{\partial t}(u_n,t_n)  + h\dfrac{\partial f}{\partial u}(u_n,t_n) \sum_{j=1}^i \beta_{ij}k_j , 
    \end{aligned} \\
    &u_{n+1} = u_n + \sum_{j=1}^s \delta_jk_j,
\end{align}
where $u_n$ is the solution at the current time step $t_n$, $u_{n+1}$ is the solution at the next time step $t_{n+1}$, $h$ is the time step ($t_{n+1}=t_n+h$), $\alpha_{ij}, \beta_{ij}, \delta_j$ are the coefficients, and 
\begin{gather}
    \alpha_i = \sum_{j=1}^{i-1} \alpha_{ij}, \\
    \beta_i = \sum_{j=1}^{i} \beta_{ij}.
\end{gather}

The Rosenbrock-type methods are ideally suited for GPU compilation because they are devoid of the typical Newton's method performed per step in stiff ODE integrators \cite{hairer1991solving}. The Newton's method requires multiple linear solves and is computationally expensive due to repeated Jacobian calculation due to no reuse of previous matrix factorizations. Rosenbrock methods only require one Jacobian evaluation and have a constant number of linear solves per step, where the matrix factorization can be cached to achieve $\mathcal{O}(N^2)$ computational cost of the linear solves, where $N$ is the dimension of the ODE. Since GPUs are efficient in performing multiple small tasks in parallel,  Rosenbrock methods are expected to achieve massive speedups on GPUs in ensemble simulations.


\subsubsection{Diagonally Implicit Runge-Kutta Methods}
Another class of algorithms considered in this work is the Diagonally Implicit Runge-Kutta (DIRK) methods. The computation of a single step is given by

\begin{align}
    &k_i = f\left(u_n + h\sum_{j=1}^{i}a_{ij}\kappa_j, t_n + c_{i}h\right), \\
    &u_{n+1} = u_n + h\sum_{i=1}^{s}b_i\kappa_i,
\end{align}
where $a_{ij}$, $b_i$, and $c_j$ are the scalar constants. The class of DIRK methods considered in our benchmarking is the Explicit Singly Diagonal Implicit Runge-Kutta (ESDIRK) methods. ESDIRK methods are characterized by $a_{11} = 0$ and $a_{i,i} = \mu$, for some constant $\mu$. Particularly, we are interested in Kvaerno methods, a class of A-L stable stiffly-accurate ESDIRK method \cite{kvaerno2004singly}. The methods are suitable for benchmarking comparison as these are only available methods in other open-source software such as in Diffrax \cite{kidger2021on}.

\subsection{Stochastic Differential Equations (SDEs)} \label{section: sdes}

SDEs are extensions of ODEs which include inherent randomness. SDEs are used as models in many domains such as quantitative finance \cite{black1973pricing,merton1973theory}, systems biology \cite{wilkinson2018stochastic}, and simulation of chemical reaction networks \cite{gillespie2000chemical}. SDEs are formally defined as
\begin{equation}
    dX_t = a(X_t,t)dt + b(X_t,t)dW_t,
\end{equation}
where $W_t$ is a wiener process and $dW_t$ is a Gaussian random variable $W_{t+dt} - W_t \sim \mathcal{N}(0, dt)$, where $\mathcal{N}(0, dt)$ is the normal distribution with zero mean and variance $dt$ \cite{kloeden1992stochastic}. The general interest is in ensembles of SDE solutions for the purpose of calculating moments or simple analytical functions of moments, such as the mean and variance of the solution, and thus SDEs are a particularly strong application for ensemble parallelization of the solution process. For this reason, methods which have accelerated convergence for the calculation of the moments, known as high weak order solvers, have gained traction in the literature as a potentially performant method for numerically analyzing such solutions. One such class of methods are the stochastic explicit "$s$" stage Runge-Kutta methods:
\begin{multline*}
    \eta_j = \Tilde{X}_{t} + h\sum_{j=1}^s\lambda_{ji}a(\eta_j, t + \mu_jh) + \sum_{k=1}^m\Delta\Tilde{W}_n^k\sum_{j=1}^s\lambda^k_{ji}b(\eta_j,t+\mu_jh), \ j = 1,\hdots,s,
\end{multline*}
\begin{multline*}
    \Tilde{X}_{t+h} = \Tilde{X}_{t} + h\sum_{j=1}^s\alpha_ja(\eta_j, t + \mu_jh) + \sum_{k=1}^m\Delta\Tilde{W}_n^k\sum_{j=1}^s\beta^k_jb(\eta_j,t+\mu_jh) + R,
\end{multline*}
where $\alpha_j,\beta_j^k, \mu_j, \lambda_{ij}, \gamma_{ij}^k$ are the constants that define the particular stochastic Runge-Kutta method and $R$ is the fit term \cite{kloeden1992stochastic, tocino2002runge}. Adaptive time-stepping techniques using similar rejection sampling and PI-controller approaches to ODEs have been adapted to SDE solver software \cite{rackauckas2020stability}. 

\section{The GPU ecosystem in Julia and cross-platform GPGPU programming}

Using high-level languages to program hardware accelerators traditionally means either using a library approach or a domain-specific-language (DSL) approach. The library approach focuses on providing array abstractions to call optimized high-level operators written and optimized in another language. Prime examples of this approach are ArrayFire~\cite{malcolm2012arrayfire} and CuNumpy~\cite{cupy_learningsys2017}. Often
these systems provide some mechanism of user extendability, but often it is in terms of the underlying system language and not the host language. DSL approaches, such as JAX~\cite{jax2018github}, embed a new language into the host language that provides domain-specific concepts and limits expressibility to a subset of operations that are representable by the DSL. This requires the user to rewrite their application in ways that are compatible with the DSL. 

Compiling a high-level language directly to hardware accelerators like GPUs is challenging because these languages often rely on accessing the run-time library, managed memory and garbage collection, interpreted execution, and other constructs that are difficult to use on GPUs or are even in conflict with the hardware design. Numba~\cite{Lam_Pitrou_Seibert_2015} and JuliaGPU~\cite{besard2018effective} retarget a subset of the language for execution on hardware accelerators. In contrast to Numba, which is a reimplementation of Python, JuliaGPU repurposes Julia's existing CPU-oriented compiler for the purpose of generating code for GPUs.

Over time this has allowed the subset of the language that is directly executable on the GPU to grow and provide the basis for an effective, performant, and highly accessible programming model for GPUs.
This model spans from low-level GPU kernel programming with direct access to advanced hardware features to the high-level array abstractions~\cite{besard2019rapid} provided by Julia.

\subsection{Supporting multiple GPU platforms}

Originally JuliaGPU only supported hardware accelerators by NVIDIA (CUDA). As hypothesized~\cite{besard2018effective}, the same approach could be extended to target other hardware platforms. Instead of re-implementing a full-fledged compiler for each new platform, the common infrastructure pieces were abstracted into a single unified compiler interface GPUCompiler.jl~\cite{gpucompilerjl} and a unified array interface GPUArrays.jl~\cite{gpuarraysjl}. Despite its name, GPUCompiler.jl is not limited to only GPU platforms and is also used to target non-GPU accelerators and specific CPU platforms.

With GPUCompiler.jl offering reusable functionality to configure the Julia compiler and LLVM providing the ability to generate high-quality machine code, Julia is well-positioned to target different accelerator platforms. Most major GPU platforms are supported: CUDA.jl~\cite{besard2018effective} for NVIDIA GPUs using the CUDA toolkit, AMDGPU.jl~\cite{julian_samaroo_2023_7641665} for AMD GPUs through ROCm, oneAPI.jl \cite{besard_tim_2023_7789142} for Intel GPUs with oneAPI, and Metal.jl~\cite{besard_tim_2023_7789146} for Apple M-series GPUs with the Metal libraries. These backends are relatively simple and can be developed and maintained by small teams. Meanwhile, other languages and frameworks often struggle to provide native support for all but the most popular platforms. In the case of Python, for example, adding a Numba backend involves significant effort, and as such, Apple GPUs are not yet supported.\footnote{\href{https://github.com/numba/numba/issues/5706}{https://github.com/numba/numba/issues/5706}} Similarly, as of May 2023, JAX does not support AMD \footnote{\href{https://github.com/google/jax/issues/2012}{https://github.com/google/jax/issues/2012}} or Apple GPUs \footnote{\href{https://github.com/google/jax/issues/8074}{https://github.com/google/jax/issues/8074}}, because it requires special support in the Accelerated Linear Algebra (XLA) compiler. Unless there is sizable traction, scientific computing with these languages is not able to leverage different GPU vendors where that could have been beneficial for, e.g., HPC and AI/ML workloads~\cite{obenschain20, brown2020design}.

To facilitate working with multiple GPU platforms, Julia offers a powerful array abstraction that makes it possible to write generic code. The abstractions are implemented by each backend, either using native kernels or by reusing existing functionality. For performance reasons, common operations such as matrix multiplication are implemented by dispatching to vendor-specific libraries such as CUBLAS for NVIDIA GPUs and Metal's Performance Shaders for Apple GPUs. Higher-order operations such as \texttt{map}, \texttt{broadcast}, and \texttt{reduce} are implemented using native kernels. This makes it possible to compose them with user code, often obviating the need for custom kernels. Work by Besard et al. \cite{besard2019rapid} has shown that this makes it possible to quickly prototype code for multiple platforms while achieving good performance. To achieve maximum performance, important operations can still be specialized using custom kernels that are optimized for the platform at hand and include application-specific knowledge.

\subsection{Abstractions for kernel programming}

Even though Julia supports multiple GPU platforms, it can quickly become cumbersome to write compatible kernels for each one. For example, kernels need to adhere to specific device APIs that are offered by the platform. With a variety of device backends available, a programmer's dream is to write one kernel that can be instantiated and launched for any device backend without modifications of the higher-level code, all without sacrificing performance. In Julia, this is possible with the KernelAbstractions.jl~\cite{valentin_churavy_2023_7770454} package, which provides a macro-based dialect that hides the intricacies of vendor-specific GPU programming. Kernels can then be instantiated for different hardware accelerators, including CPUs and GPUs. For GPUs, full support is provided for NVIDIA (CUDA), AMD (ROCm), Intel (oneAPI), and Apple (Metal).
For the GPU ODE solvers presented in this article, KernelAbstractions.jl is used to target these different backends, essentially from the same high-level kernel code.

The development of the GPU ecosystem was not a separate effort rather, DiffEqGPU.jl has been one of the driving projects of the Julia GPU ecosystem since the inception of the new approaches (GPUCompiler.jl and KernelAbstractions.jl) in 2019. As such, the developers of those tools are co-authors of this work, as this is the first application that displays the full vendor-agnostic feature set that the tooling aims to provide.

\section{Massively Data-Parallel GPU Solving of Independent ODE Systems}

\begin{figure}
  \centering
  \includegraphics[width=0.5\linewidth]{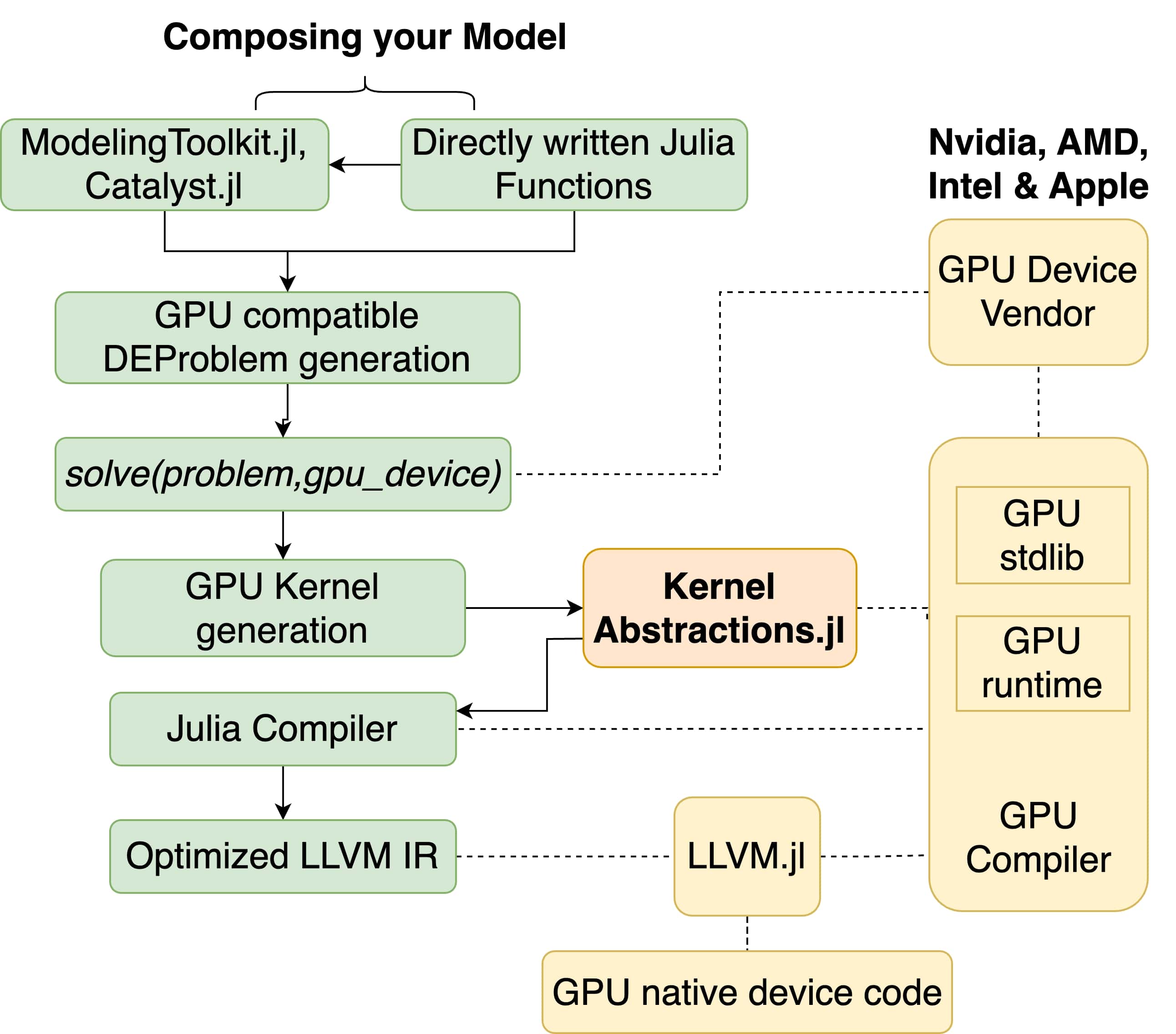}
  \caption{Overview of the automated translating and solving of differential equations for GPUs for massively data-parallel problems. The solid lines indicate the code flow, whereas the dashed lines indicate the extension interactions.}
  \label{fig:gpu_solve_flowchart}
\end{figure}

This article considers two approaches for parallelizing ensemble problems on GPUs, both of them automatically translating and compiling the differential equation. The first approach is easily extensible, compatible with any existing solver, and relies on GPU vectorization. This approach is similar to the other high-level software we described in the introduction, and we will show that this approach is not performance optimal and has significant overheads. The second strategy reduces this overhead by generating custom GPU kernels, requiring numerical methods to be programmed within it. A brief overview of the automation is depicted in Figure \ref{fig:gpu_solve_flowchart}. Both of the programs are composable with Julia's SciML \cite{Rackauckas_2017_Julia} ecosystem, where users can write models compatible with standard SciML tools such as DifferentialEquations.jl, and DiffEqGPU.jl will automatically generate the functions which can be invoked from within a GPU kernel. Moreover, SciML is composed of polyglot tools allowing use of its libraries from other languages such as R, allowing even the use of our GPU-accelerated solvers from other programming languages.\footnote{\href{https://cran.r-project.org/web/packages/diffeqr/vignettes/gpu.html}{https://cran.r-project.org/web/packages/diffeqr/vignettes/gpu.html}}

\subsection{\texttt{EnsembleGPUArray}: Accelerating Ensemble ODEs with GPU Array Parallelism} \label{EnsembleGPUArray}

\subsubsection{Identifying parallelism and problem construction}
\label{sec:problem_construction}

For an ODE with $n$ states, $m$ parameters, and $N$ required simulations with different parameters, there exist $n\times N$ states to keep track of. This problem can be formulated as solving one ODE
\begin{equation}
    \frac{dU}{dt} = F(U,P,t)
\end{equation}
where
\begin{equation}
U = 
\begin{bmatrix}
    u_{11} & u_{12} & \cdots & u_{1N} \\
    u_{21} & u_{22} & \cdots & u_{2N} \\
    \vdots & \vdots & \ddots & \vdots \\
    u_{n1} & u_{n2} & \cdots & u_{nN} \\
\end{bmatrix}_{n \times N},
\end{equation}
\begin{equation}
P = 
\begin{bmatrix}
    p_{11} & p_{12} & \cdots & p_{1N} \\
    p_{21} & p_{22} & \cdots & p_{2N} \\
    \vdots & \vdots & \ddots & \vdots \\
    p_{m1} & p_{m2} & \cdots & p_{mN} \\
\end{bmatrix}_{m \times N}, 
\end{equation}
\begin{equation}
    F = \begin{bmatrix}
    f(u,p_{1:m,1},t) & \cdots & f(u,p_{1:m,N},t) \\
\end{bmatrix}_{n\times N},
\end{equation}
and $p_{1:m,j}$ denotes the $j^\mathrm{th}$ column of the $P$ matrix. In this form, we can parallelize the computation over GPU threads, where each thread only accesses and updates the column of $U$ in parallel. This allows the computation of the quantities which depend on $U$ to happen in parallel. When solving ODEs, these quantities are generally the right-hand side (RHS) of ODE $f$, the Jacobian $J$, and even the event handling (callbacks). We perform these array-based computations by calling the functions within custom-written GPU kernels, updating each column of the $U$ asynchronously.

\subsubsection{Translating ODE solves over GPU using KernelAbstractions.jl}

In order to fully leverage the GPU ecosystem in Julia, several changes were made to it to allow DiffEqGPU.jl to have seamless performance and composability. Changes such as the generalization of SciML kernels from CUDA-specific to backend-agnostic, launch parameters tuning, and delivering usability of newer backends such as Apple metal are some of the initiatives to make this work a reality, which in turn is helping the broader community. Hence, our work provides a roadmap for how others can achieve true vendor agnosticism and the reality of the work necessary.

The GPU kernels are written using KernelAbstractions.jl~\cite{valentin_churavy_2023_7770454}, which allows for the instantiation of the GPU kernels for multiple backends. KernelAbstractions.jl performs a limited form of auto-tuning by optimizing the launch parameters for occupancy. Since these kernels have a high residency, preferring a launch across many blocks has been shown to be beneficial. We instantiate the kernels with the problem defined as normal Julia functions that the kernel is specialized upon. Using a Just-In-Time (JIT) compilation approach, we thus generate a new kernel where the solver and the problem definition are co-optimized.

\begin{figure}
  \centering
  \includegraphics[width=0.5\linewidth]{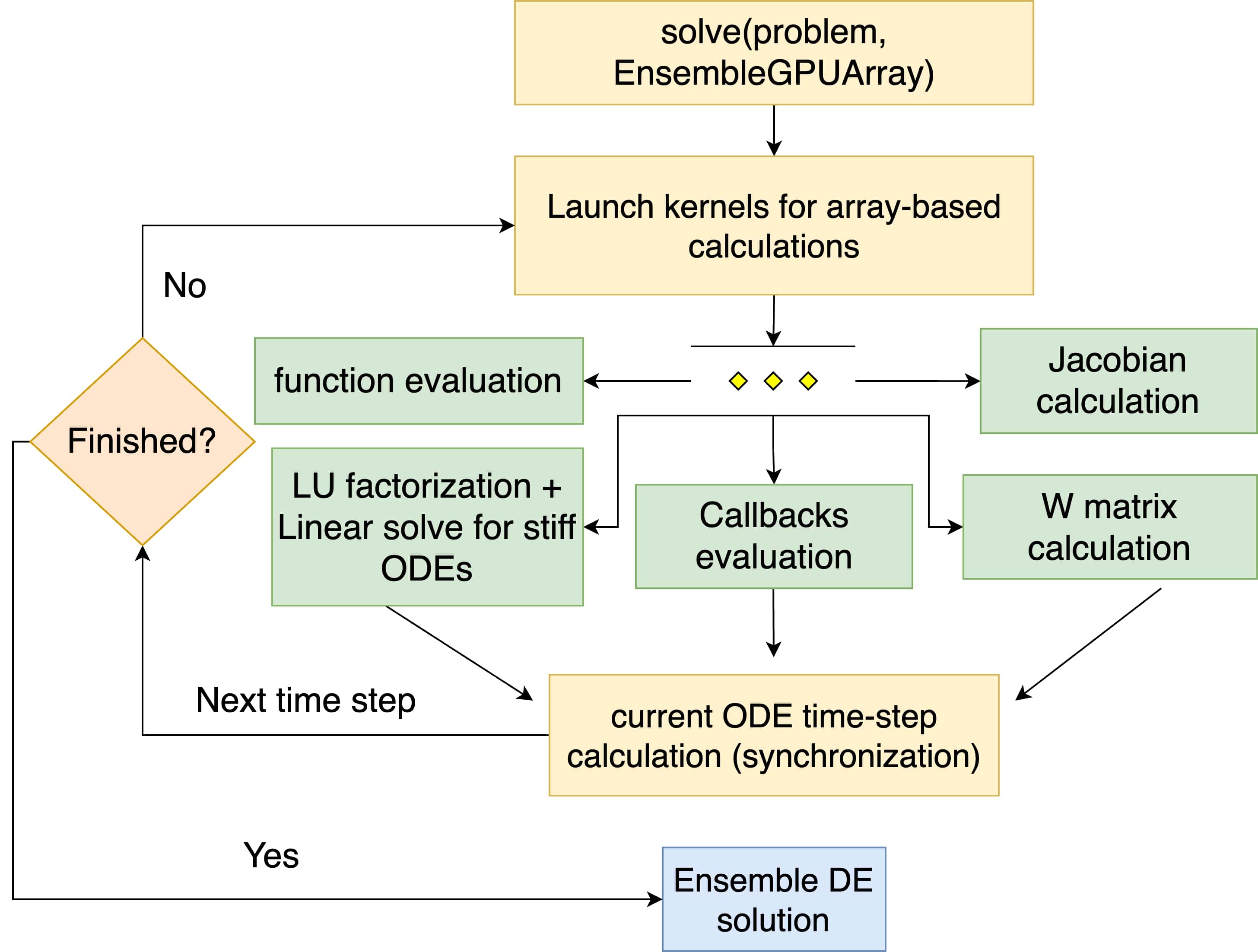}
  \caption{The EnsembleGPUArray flowchart.}
  \label{fig:ensemblegpuarray_solve_flowchart}
\end{figure}

After calculating the dependents on $U$ is completed, synchronization is required to calculate the next step of the integration. \texttt{EnsembleGPUArray} essentially parallelizes the operation involving the state $U$ within the single time step of the ODE integration. This simple approach allows composability and easy integration with the vast collection of numerical integration solvers in DifferentialEquations.jl \cite{Rackauckas_2017_Julia}. An option to simultaneously offload a subset of the solutions to the CPUs provides additional flexibility to the user to leverage the CPU cores. Moreover, users can take advantage of the multiple GPUs over clusters to perform the simulations of the ensemble problems via this tutorial.\footnote{\href{https://docs.sciml.ai/DiffEqGPU/dev/tutorials/multigpu/}{https://docs.sciml.ai/DiffEqGPU/dev/tutorials/multigpu/}} Figure \ref{fig:ensemblegpuarray_solve_flowchart} is an overview of the process.

\subsubsection{Batched LU: Accelerating Ensemble of Stiff ODEs}

Stiff ODE solvers require repeatedly solving the linear system $W^{-1} b$ where $W = -\gamma I + J$, $\gamma$ is a constant real number, and $J$ is the Jacobian matrix of the RHS of the ODE. The $W$ matrix of the batched ODE problem in Section \ref{sec:problem_construction} has a block diagonal structure:
\begin{equation}
W = 
\begin{bmatrix}
    -\gamma I + J_{1} &        &   &  \\
           & -\gamma I + J_{2} &  & \\
     &  & \ddots & \\
           &  & & -\gamma I + J_{N} \\
\end{bmatrix},
\end{equation}
where $(J_{k})_{i j} = \frac{{\partial f}_i}{{\partial u}_{j k}}$. The block diagonal system can be efficiently solved by computing the LU factorization, and forward, and backward substitutions of each block of $W$ in the GPU kernel.

\subsubsection{Drawbacks of the Array Ensemble Approach}

The main drawback of this approach is that each array operation inside of the ODE solver requires a separate GPU kernel launch. However, in explicit Runge-Kutta methods as described in Sections \ref{section: odes} and \ref{section: sdes}, most of the operations are linear combinations and column-wise parallel applications of the ODE model $f$, and are thus $\mathcal{O}(N)$ operations. Array-based GPU DSLs are typically designed to be used with $\mathcal{O}(N^3)$ operations which are common in neural network applications (such as matrix multiplication) in order to more easily saturate the kernels to overcome the overhead of kernel launch. While the ODE solvers are written in a form that automatically fuses the linear combinations to reduce the total number of kernel calls, thus reducing the overall cost \cite{wang2010kernel}, we will see in the later benchmarks (Section \ref{section:cpu_benchmarks}) that each of the array-ensemble GPU ODE solvers has a high fixed cost due to the total overhead of kernel launching.

In addition, the parallel array computations of each step of the solver method need to be completed before proceeding to the next time step of the integration. Adaptive time-stepping in ODEs allows variable time steps according to the local variation in the ODE integration, allowing optimal time-stepping. Trivially, the ODE can have different time-stepping behavior for other parameters, as they form part of the "forcing" function $f(u,p,t)$. The implicit synchronization of the parallel computations necessitates the same time-stepping for all the trajectories by virtue of solving all trajectories as a single ODE. 

\subsection{\texttt{EnsembleGPUKernel}: Accelerating Ensemble of ODEs with specialized kernel generation for entire ODE integration}

\begin{figure}
  \centering
  \includegraphics[width=0.5\linewidth]{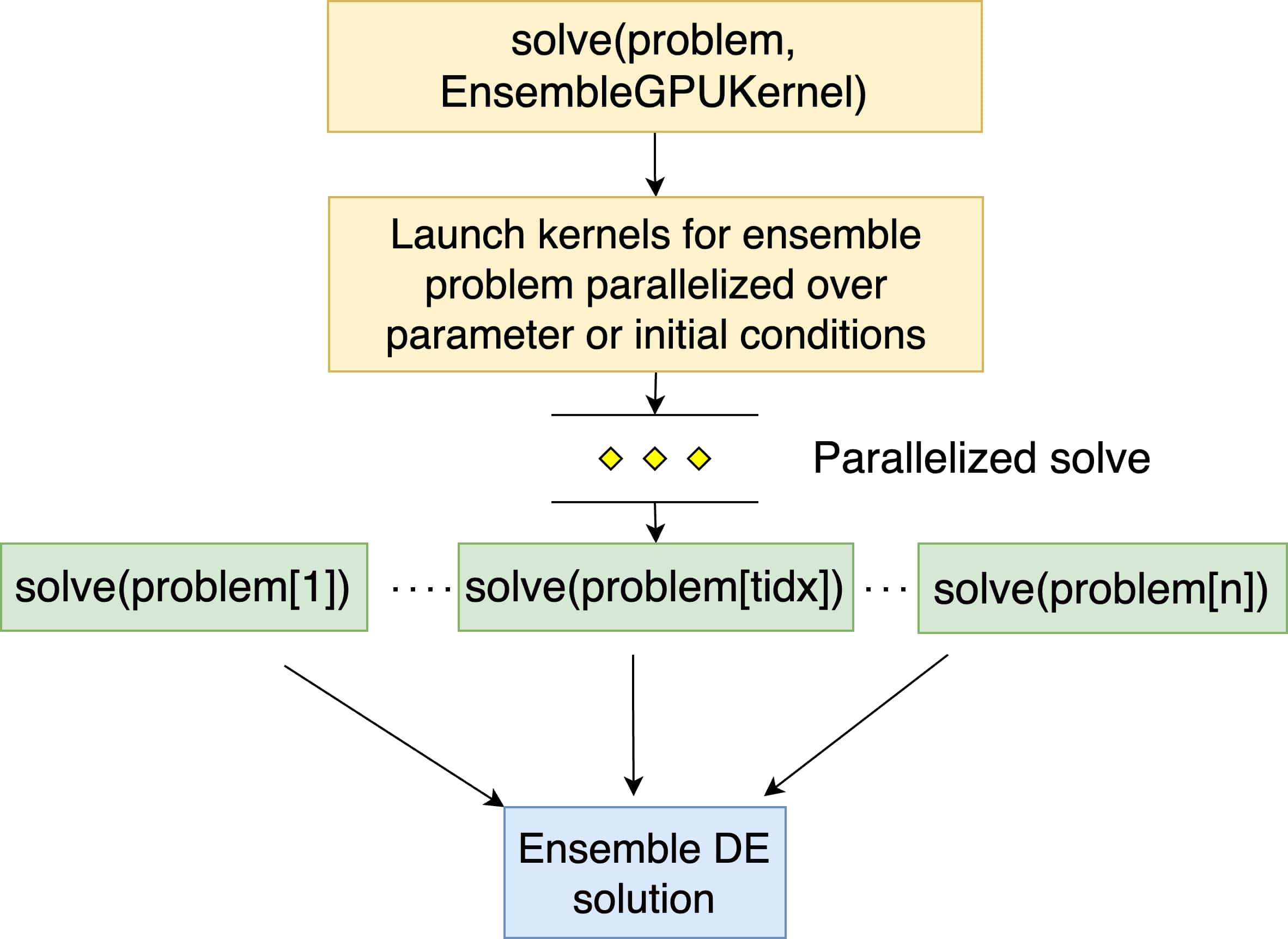}
  \caption{The EnsembleGPUKernel flowchart.}
  \label{fig:ensemblegpukernel_solve_flowchart}
\end{figure}

The \texttt{EnsembleGPUArray} requires multiple kernel launches within a time step, which causes large overheads due to the numerous load and storage operations to global memory. In order to completely eliminate the overhead of kernel launches, a separate implementation denoted  \texttt{EnsembleGPUKernel} generates a single model-specific kernel for the full ODE integration. For stiff ODEs, the Jacobian can be calculated with Automatic Differentiation (AD) invoked within the kernels. Each thread accesses the data-augmented ODE to analyze, and the solving of all the ODEs is completely asynchronous. The process is briefly outlined in Figure \ref{fig:ensemblegpukernel_solve_flowchart} and an example is given in Listing \ref{listing:EnsembleGPUKernel_Example}.

\begin{lstlisting}[caption={Example of the kernel performing ODE integration},captionpos=b, label = {listing:EnsembleGPUKernel_Example}]

    @kernel function tsit5_kernel(@Const(probs), _us, _ts, dt)
        # Get the thread index
        i = @index(Global, Linear)
        # get the problem for this thread
        prob = @inbounds probs[i]
        # get the input/output arrays for this thread
        ts = @inbounds view(_ts, :, i)
        us = @inbounds view(_us, :, i)
        # Setting up initial conditions and integrator
        integrator = init(...)
        # Perform ODE integration until completion
        while cur_time < final_time
          step!(integrator, ts, us)
          savevalues!(integrator, ts, us)
        end
        # Perform post-processing
        ...
    end
\end{lstlisting}

The approach described seems deceptively simple but requires clever maneuvers to successfully compile the kernel on GPU. Allocating arrays within GPU kernels is not possible, as Julia's CUDA.jl does not support dynamic memory allocation on the GPU. However, solving ODEs requires storing intermediate computations, normally using array allocations. The vast features of DifferentialEquations.jl rely on operations such as broadcast operations, dynamic allocations, and dynamic function invocation -- most of which are GPU incompatible.
The solution is to fully stack allocate all intermediate arrays and to perform the ODE integration within a custom GPU kernel implementing the numerical integration procedure.
This restricts the user to the set of the already defined ODE solvers in the package and requires simple versions of the ODE solvers to be manually written as GPU kernels.

\subsubsection{Kernel-Specialized ODE Solvers}

The ODE solvers that are currently available with \texttt{EnsembleGPUKernel} are:
\begin{itemize}
    \item \texttt{GPUTsit5}: A custom GPU-kernelized implementation of Tsitouras' $5^\mathrm{th}$-order solver, a Runge-Kutta 5(4) order method \cite{tsitouras2011runge}. Serves as a go-to choice for solving non-stiff ODEs. Performs well for medium to high tolerances. More efficient and precise than the popular Dormand-Price 5(4) \cite{hairer1993solving} Runge-Kutta pair,  which is a common default solver choice in packages such as MATLAB \cite{Shampine_1997_MATLAB} and torchdiffeq \cite{chen2018neural}. Has free $4^\mathrm{th}$-order interpolation support.
    \item \texttt{GPUVern7}: A custom GPU-kernelized implementation of Verner's $7^\mathrm{th}$-order solver, a Runge-Kutta 7(6) order method \cite{verner2010numerically}. Performs best at medium and low tolerances. Has a $7^\mathrm{th}$-order lazy interpolation scheme.
    \item \texttt{GPUVern9}: A custom GPU-kernelized implementation of Verner's $9^\mathrm{th}$-order solver, a Runge-Kutta 9(8) order pair \cite{verner2010numerically}. Performs best at extremely low tolerances. Has a $9^\mathrm{th}$-order lazy interpolation scheme.
    \item \texttt{GPURosenbrock23}: A custom GPU-kernelized implementation of an order 2/3 L-Stable Rosenbrock-W method \cite{Rosenbrock1963,hairer1991solving,Shampine_1997_MATLAB}. Is good for very stiff equations with oscillations at high tolerances. Employs a second-order stiff-aware interpolation. Also supports Differential Algebraic Equations (DAEs) in mass-matrix form, i.e., $M\frac{du}{dt} = f(u,p,t)$, where $M$ is the mass matrix. 
    \item \texttt{GPURodas4}: A custom GPU-kernelized implementation of a fourth-order A-stable stiffly stable Rosenbrock method \cite{Rosenbrock1963,hairer1991solving}. Is suitable for problems at medium tolerances. Employs a third-order stiff-aware interpolation.
    \item \texttt{GPURodas5P}: A custom GPU-kernelized implementation of a fifth-order A-stable stiffly stable Rosenbrock method \cite{Rosenbrock1963,Steinebach_2023_Rodas5P}. Is suitable for problems at medium tolerances. Employs a fourth-order stiff-aware interpolation which has better stability in the adaptive time-stepping embedding.
\end{itemize}
These choices are based on the SciMLBenchmarks, which has extensive comparisons between ODE solvers.\footnote{\href{https://docs.sciml.ai/SciMLBenchmarksOutput/stable/}{https://docs.sciml.ai/SciMLBenchmarksOutput/stable/}} For stiff ODEs, the Rosenbrock-type methods are ideally suited for GPU compilation because they are devoid of the typical Newton's method performed per step in stiff ODE integrators \cite{hairer1991solving}. On CPU, implicit ODE solvers achieve top performance by using the property that the inverted Jacobian of Newton's method does not need to be exact, and thus efficient integrators such as CVODE \cite{Hindmarsh_2005} and those in DifferentialEquations.jl adaptively reuse the same Jacobian factorization from multiple time steps. For sufficiently large equations, this trades the $\mathcal{O}(N^3)$ factorization operation for more iterations of Newton's method with the same inverse factor and thus $\mathcal{O}(N^2)$ linear solves. However, these properties do not transfer well to the GPU setting since (a) branching within a GPU warp is resolved such that every concurrent solve requires the same number of iterations, thus leading to more linear solves than necessary for most of the systems at each step and (b) the systems are sufficiently small so that the factorization is not the dominating factor in the compute cost. 

Rosenbrock methods only require one Jacobian evaluation and have a constant number of linear solves per step, where the matrix factorization can be cached to a single factorization per time step. This more static integration procedure is also compensated by higher stage order (i.e., convergence rate on highly stiff ODEs and DAEs) and smaller leading truncation error coefficients, effectively leading to less steps being required to reach the same error. For this reason, Rosenbrock methods are the most efficient solver on CPU for sufficiently small ODEs (approximately less than 20), competitive until the LU-factorization cost becomes dominant (around 100 ODEs). Some prior research has shown results with ODE extrapolation methods where the LU factorization does not have the dominating cost and hence no suitable gains from its parallelization other than parallelizing multiple computations of LU  \cite{utkarsh2022parallelizing}. 

Integration of automatic differentiation: Automatic differentiation is a way to compute exact derivatives of mathematical computer programs \cite{corliss2002automatic}. With the change to high-order Rosenbrock methods, having accurate Jacobians (non-finite difference) is a necessary requirement for Rosenbrock methods to achieve high-order convergence \cite{hairer1991solving}. Thus, while previous LSODA approaches could get away with finite difference approximations through the relaxation in nonlinear solver steps, our approach, which would be more GPU performant due to the greatly reduced branching, requires mixing this automatic differentiation to achieve full accuracy and performance. Integration of solvers is done with forward-mode automatic differentiation within a GPU kernel. In particular, we extended the LU factorization in the nonlinear solve step to statically compile seamlessly for arbitrary size of the ODE.

Given the extra advantages of the GPU and the fact that the embarrassingly parallel context is limited to this size of systems, we will see that Rosenbrock methods are outperformed by the Newton-based stiff ODE solvers quite handily. 

\subsubsection{Kernel-Specialized SDE solvers}

Currently, DiffEqGPU.jl only supports fixed time-stepping in SDEs with \texttt{EnsembleGPUKernel}.
\begin{itemize}
    \item \texttt{GPUEM}: A custom GPU-kernelized fixed time-step implementation of Euler-Maruyama method \cite{kloeden1992stochastic}. Supports diagonal and non-diagonal noise.
    \item \texttt{GPUSIEA}: A custom GPU-kernelized fixed time-step implementation of weak order 2.0 \cite{tocino2002weak}, stochastic generalization of midpoint method. Supports only diagonal noise.
\end{itemize}

\section{Benchmarks and Case studies}

\subsection{Setup} \label{section:setup}

To compare different available open-source programs with GPU-accelerated ODE solvers, we benchmark them with several NVIDIA GPUs: one being a typical compute node GPU, Tesla V100, and the other being a high-end desktop GPU, Quadro RTX 5000. Performance comparison of the kernel-based ODE solvers with different GPU vendors is also carried out. Except for Apple having an integrated GPU, we use dedicated desktop GPUs. For NVIDIA, we benchmark on Quadro RTX 5000 (11.15 TFLOPS), Vega 64 for AMD (10.54 TFLOPS), A770 (19.66 TFLOPS) for Intel, and M1 Max (10.4 TFLOPS) for Apple. The DE problems involve single precision (Float32) on GPUs. The CPU benchmarks are executed using double precision (Float64), and are timed on an Intel Xeon Gold 6248 CPU @ 2.50GHz with 16 enabled threads. Using double precision on CPUs is faster than the single precision for our use-case and processor.

To facilitate the seamless transition from CPU to GPU without requiring any modifications to the original code, \texttt{EnsembleGPUKernel} was developed to mimic the SciML ensemble interface. Nevertheless, this GPU-aspect-hiding approach always passes the problem to the GPU and returns the result to the CPU, reducing overall performance. To avoid conversion overheads and for a fair comparison among other software, we developed a lower-level API\footnote{\href{https://docs.sciml.ai/DiffEqGPU/dev/tutorials/lower_level_api/}{https://docs.sciml.ai/DiffEqGPU/dev/tutorials/lower\_level\_api/}} that closely resembles other APIs. The timings for each software only report time spent solving the ensemble ODE. The benchmarking literature discusses the topic of reporting times for benchmarking and uses the minimum as an approximation of the ideal value, the least noisy measurement. Hence, the timings for the programs written in Julia are measured using BenchmarkTools.jl, taking the best timing. The Julia-based benchmarks were tested on DiffEqGPU.jl 1.26, CUDA.jl 4.0, oneAPI.jl 1.0, and Metal.jl 0.2.0, all using Julia 1.8. The test with AMD GPUs was done with AMDGPU.jl 0.4.8, using Julia 1.9-beta3.

The timings script for MPGOS has been borrowed from their available open-source codes.\footnote{\href{https://github.com/nnagyd/ode_solver_tests}{https://github.com/nnagyd/ode\_solver\_tests}} They have been tested with CUDA toolkit 11.6 and C++ 11. The programs are run at least ten times for JAX and PyTorch. The JAX-based programs are run on 0.4.1 with Diffrax 0.2.2. Programs based on PyTorch are tested with PyTorch nightly 2.0.0.dev20230202, and a custom installation of torchdiffeq to extend support to \texttt{vmap} is used.\footnote{\href{https://github.com/utkarsh530/torchdiffeq/tree/u/vmap}{https://github.com/utkarsh530/torchdiffeq/tree/u/vmap}}
Both programs are tested on Python 3.9. 
The complete benchmark suite can be found at \href{https://github.com/utkarsh530/GPUODEBenchmarks}{https://github.com/utkarsh530/GPUODEBenchmarks}.

\subsection{Establishing efficiency of solving ODE ensembles with GPU over CPU} \label{section:cpu_benchmarks}

\begin{figure}[ht]
  \centering
  \includegraphics[width=0.7\linewidth]{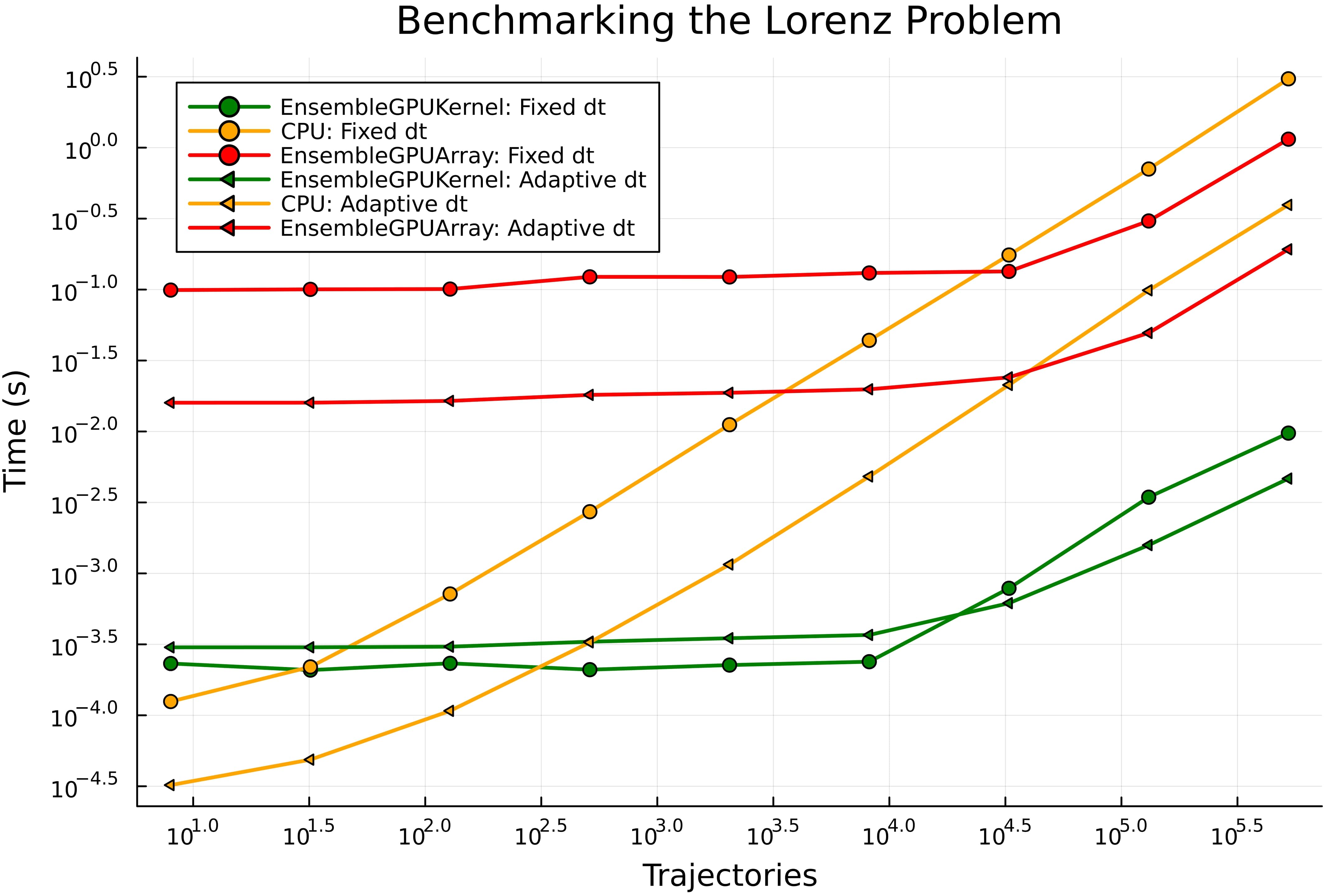}
  \caption{A comparison of time for an ODE solve for CPU vs.\ GPU. The EnsembleGPUKernel performs the best with up to $100\times$ acceleration and a lower cutoff to take advantage of parallelism.}
  \label{benchmark:cpu_lorenz}
\end{figure}

\begin{figure}[ht]
  \centering
  \includegraphics[width=0.7\linewidth]{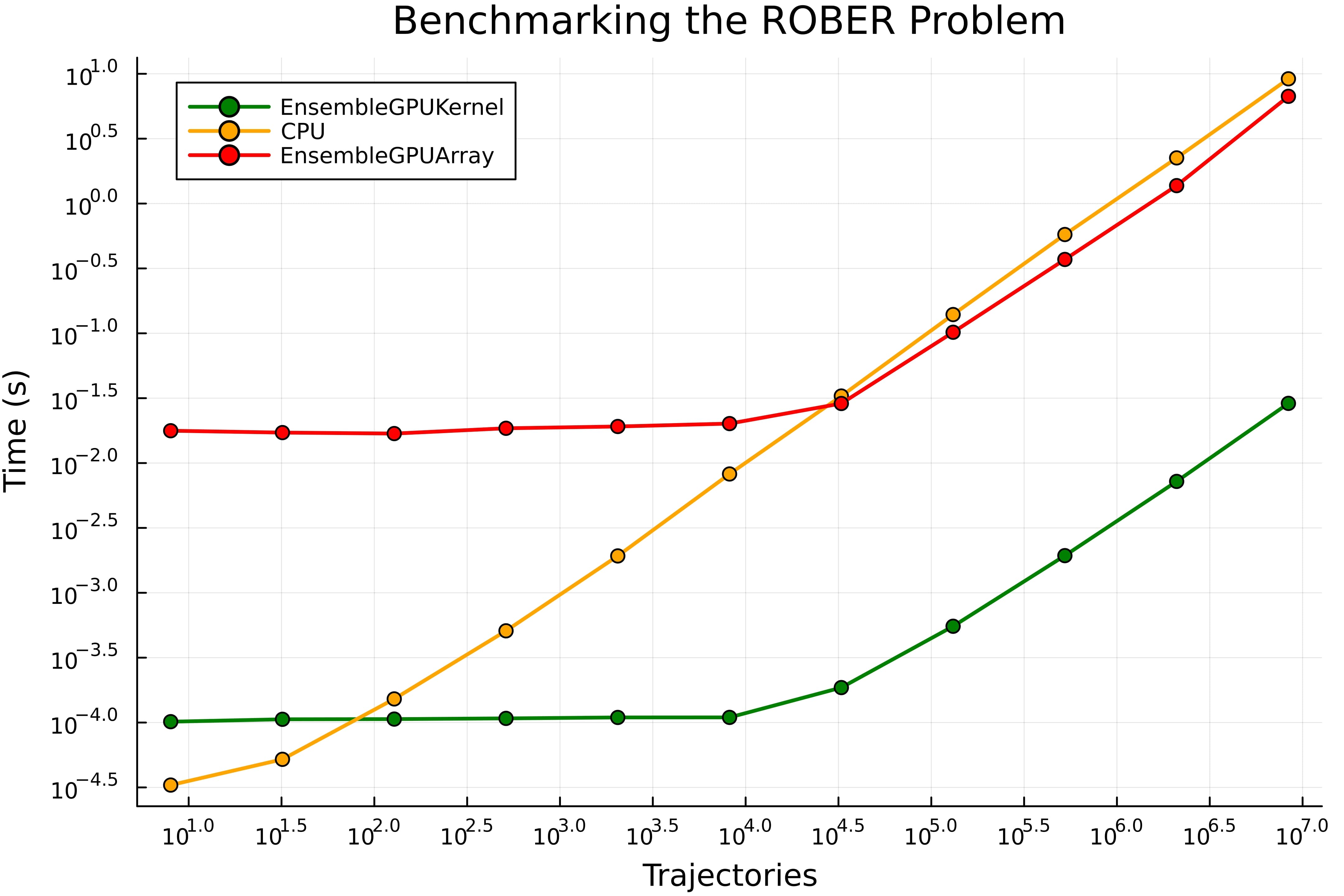}
  \caption{A comparison of the time of an ODE solve for CPU vs.\ GPU. The EnsembleGPUKernel performs the best with up to $100\times$ acceleration and a lower cutoff to take advantage of parallelism.}
  \label{benchmark:cpu_Rober}
\end{figure}

Prior researches \cite{hegedHus2021program, nagy2022art} demonstrate that solving low-dimensional ODEs over parameters and initial conditions (massively parallel) exposes superior parallelism in GPUs over CPUs. The benchmarking is performed on GPUs with single precision and CPU multithreading with double precision to take advantage of respective optimized floating point math. Indeed, the \texttt{EnsembleGPUKernel} supports our claim of lower overhead compared to \texttt{EnsembleGPUArray} and being up to $100 \times$ faster. For the benchmark problem GPU parallelism becomes superior to CPU parallelism at approximately $100$--$1000$ trajectories  (Figure \ref{benchmark:cpu_lorenz}). The solver used to benchmark the Lorenz Attractor \cite{lorenz1963deterministic} is Tsitouras' $5(4)$ Runge-Kutta method \cite{tsitouras2011runge}, both with adaptive and fixed time-stepping. Similarly, we also perform the benchmarking of stiff ODE integrators with the Robertson equation\cite{robertson1976numerical}, using the Rosenbrock23 methods \cite{Rosenbrock1963}. Figure \ref{benchmark:cpu_Rober} showcases the performance of the solvers. Table \ref{tab:bench_cpu_vs_cpu} lists the relative slowdowns of both non-stiff and stiff ODE integrators obtained using \texttt{EnsembleGPUKernel}.

\begin{table*}
    \caption{Summary of mean slowdowns of ODE integrators, benchmarking on stiff problems with different hardware \textit{(lower the better)}}
   \label{tab:bench_cpu_vs_cpu}
   \centering
    \begin{tabular}{cccl}
        \toprule
         Time-stepping &  GPU (Kernel) & GPU (Array) & CPU  \\
        \midrule
        Adaptive (Nonstiff) & 1.0$\times$ & 48.2$\times$ &  22.2$\times$ \\  
        Fixed (Nonstiff) & 1.0$\times$ & 377.6$\times$ & 110.3$\times$ \\
        Adaptive (Stiff) & 1.0$\times$ & 180.0$\times$ & 132.3$\times$ \\ 
        \bottomrule
    \end{tabular}
\end{table*}

\textbf{GPU performance at < 1000 trajectories}: Massive parallel simulations are needed to utilize GPU cores and amortize overhead costs fully. Figure \ref{benchmark:cpu_lorenz} shows that overheads are essentially constant regardless of thread count. In this range, the computations are not latency hiding; one essentially measures the kernel launch time. Hence, one would not expect superior GPU parallelism in comparison to CPU at lower trajectories because CPUs are usually free of these overheads.

\subsection{Solving billions of ODEs together: Scaling \texttt{EnsembleGPUKernel} with MPI}

The Message Passing Interface (MPI) \cite{walker1996mpi, gabriel2004open} can be directly used with Julia using MPI.jl \cite{byrne2021mpi}. Additionally, there exists support for CUDA by using a CUDA-aware MPI backend. As a testimonial of scalability, 2,147,483,648 ($\sim$2 billion) ODE solutions on the Lorenz problem were calculated using eight V100 GPU cluster nodes, in which seven GPUs amounting to 35,840 CUDA cores explicitly performed the computation of approximately 306 million ODEs. The wall-clock time of this simulation was approximately 50 seconds, which includes other latencies such as package loading, compilation, and GC times. The runtime of the MPI call (creating buffers and transferring arrays to GPUs) and solving ODEs was around 13 seconds. The runtime of only the ODE solve of 306 million trajectories was around 1.6 seconds.\footnote{These latencies could be reduced by using a GPU with larger memory or a multi-GPU per node, which was not done in our demonstration due to hardware availability.} We also note in our observations that the methods are scalable, 
and are limited by the global memory of the GPU required for storing the solutions, which can be roughly calculated by calculating the memory requirements of the number of time-points required multiplied by the different number of parameters in the simulation. Subsequently, the methods are prophesied to scale well with multiple nodes in a cluster.

\subsection{Comparison with other GPU-accelerated ODE programs} \label{section: comp_gpu_programs}

Selecting the problem for fair benchmarking on different implementations of GPU-based solvers is a task in itself, owing to the different use cases, motivations, and optimizations of these libraries. Choosing systems with low-dimensional ODEs, such as the Lorenz equation, is a suitable candidate owing to the simplicity $f(u,p,t)$ of definition, which alludes to any optimizations in calculating $f(u,p,t)$. The right-hand side solely involves additions/subtractions and multiplications, in which each floating-point operation will be interpreted as a complete FMA (Fused Multiply Accumulate) instruction. For our benchmarking purposes, $\sigma = 10.0$, $\gamma = \frac{8}{3}$, and $\rho$ in the Lorenz equation \cite{lorenz1963deterministic} is uniformly varied from $(0.0,21.0)$ generating $N$ instances of independent, parallel ODE solves, which is also the number of trajectories in DiffEqGPU.jl API. 

\begin{figure}[ht]
  \centering
  \includegraphics[width=0.7\linewidth]{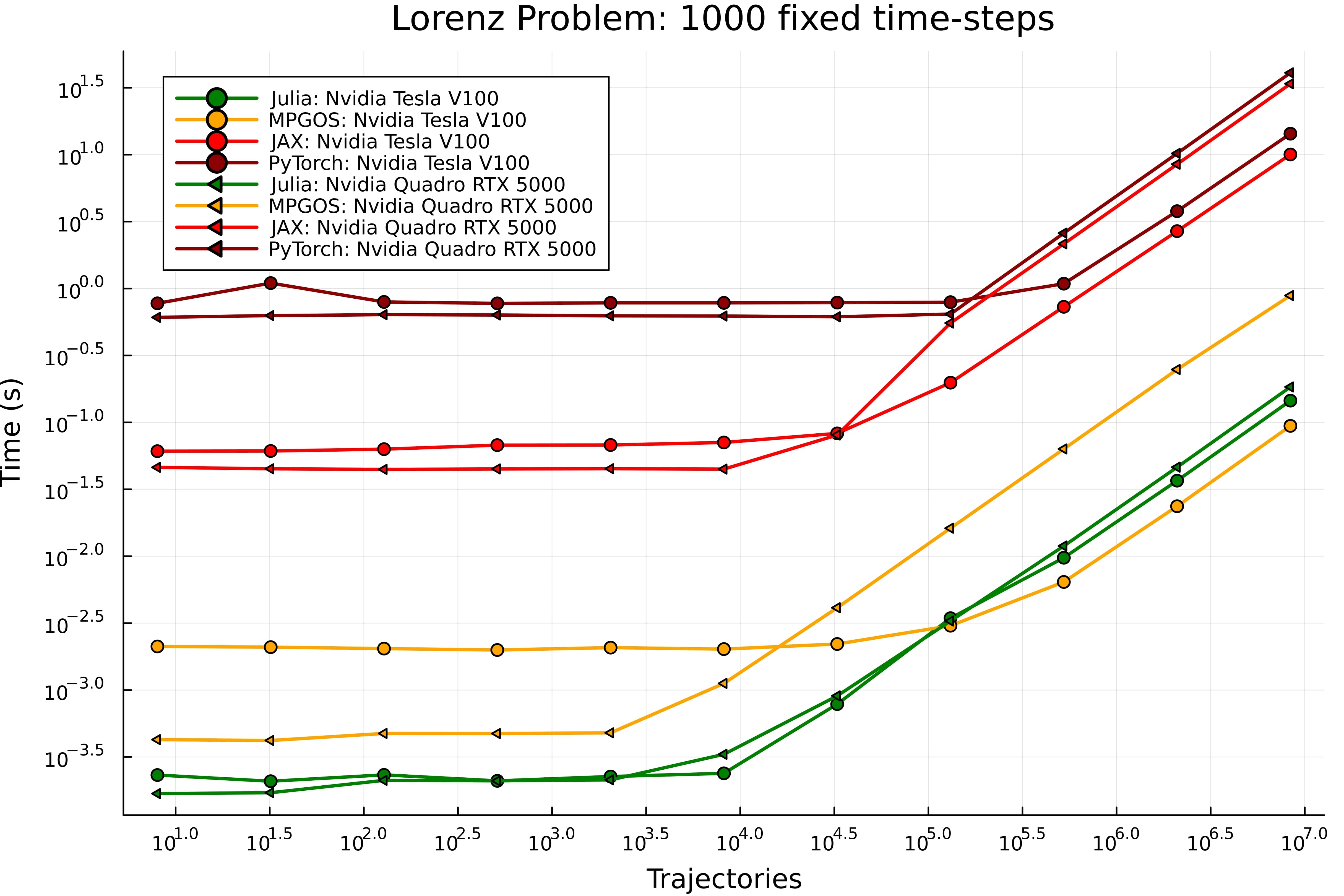}
  \caption{A comparison of the time for an ODE solve with other programs with fixed time-stepping. EnsembleGPUKernel is able to reach and sometimes outperform speed of light measure (MPGOS) and is approximately faster by 20--100$\times$ in comparison to JAX and 100--200$\times$ for PyTorch.}
  \label{benchmark:lorenz_unadaptive}
\end{figure}

\begin{figure}[ht]
  \centering
  \includegraphics[width=0.7\linewidth]{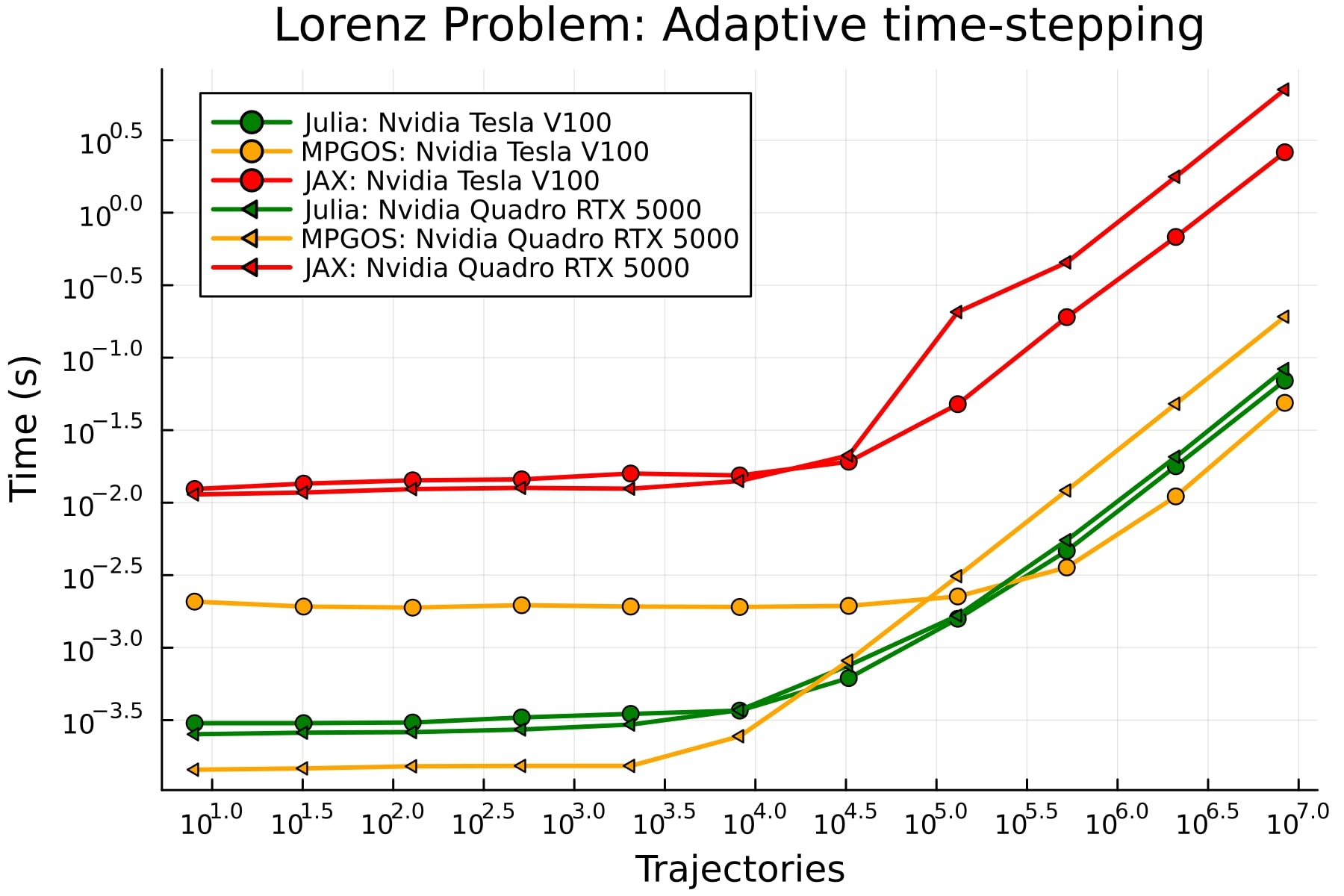}
  \caption{A comparison of the time for an ODE solve for with other programs with adaptive time-stepping. EnsembleGPUKernel is able to reach and sometimes outperform speed of light measure (MPGOS) and approximately faster by 20--100$\times$ in comparison to JAX.}
  \label{benchmark:lorenz_adaptive}
\end{figure}

The results for fixed and adaptive time-stepping are shown in separate Figures \ref{benchmark:lorenz_unadaptive} and \ref{benchmark:lorenz_adaptive} for equitable comparison. There is not any common ODE solver between these packages, so we use methods belonging to the class of $4^{\textrm{th}},5^{\textrm{th}}$ order Runge-Kutta methods, which perform similarly in the benchmarks \cite{Rackauckas_2017_Julia}. "Tsit5" was used in both Julia and JAX, "Cash-Karp" for MPGOS, and "Dopri5" for PyTorch. Fixed time-stepping implicitly assures a constant work/thread, which is ideal for GPUs. However, adaptive time-stepping within ODE integrators adjusts time steps to ensure stability and error control. This generally results in faster ODE integration times than fixed time-stepping, but may cause thread divergence as effectively different time-stepping across threads, eventually amounting to contrasting work per thread. The slowdown of different packages with respect to our implementation with different NVIDIA GPUs is listed in Tables \ref{tab:impl_comparison_rtx} and \ref{tab:impl_comparison_v100}. While our solvers are able to reach and sometimes outperform speed of light measure (C++, MPGOS), we observe a greater acceleration of approximately 20--100$\times$ in comparison to JAX and 100--200$\times$ for PyTorch, both of which rely on vectorized map style of parallelism. The authors were not able to compile adaptive time-stepping results for PyTorch, as \texttt{vmap} currently does not support all the internal operations required by the PyTorch code base. Notably, the array abstraction parallelism approach of PyTorch and JAX performs similarly to the demonstrated efficiency of \texttt{EnsembleGPUArray}, providing clear evidence that the performance difference is due to the fundamental approach itself and not due to efficiencies or inefficiencies in the implementation of the approach.

\begin{figure}[ht]
  \centering
  \includegraphics[width=0.7\linewidth]{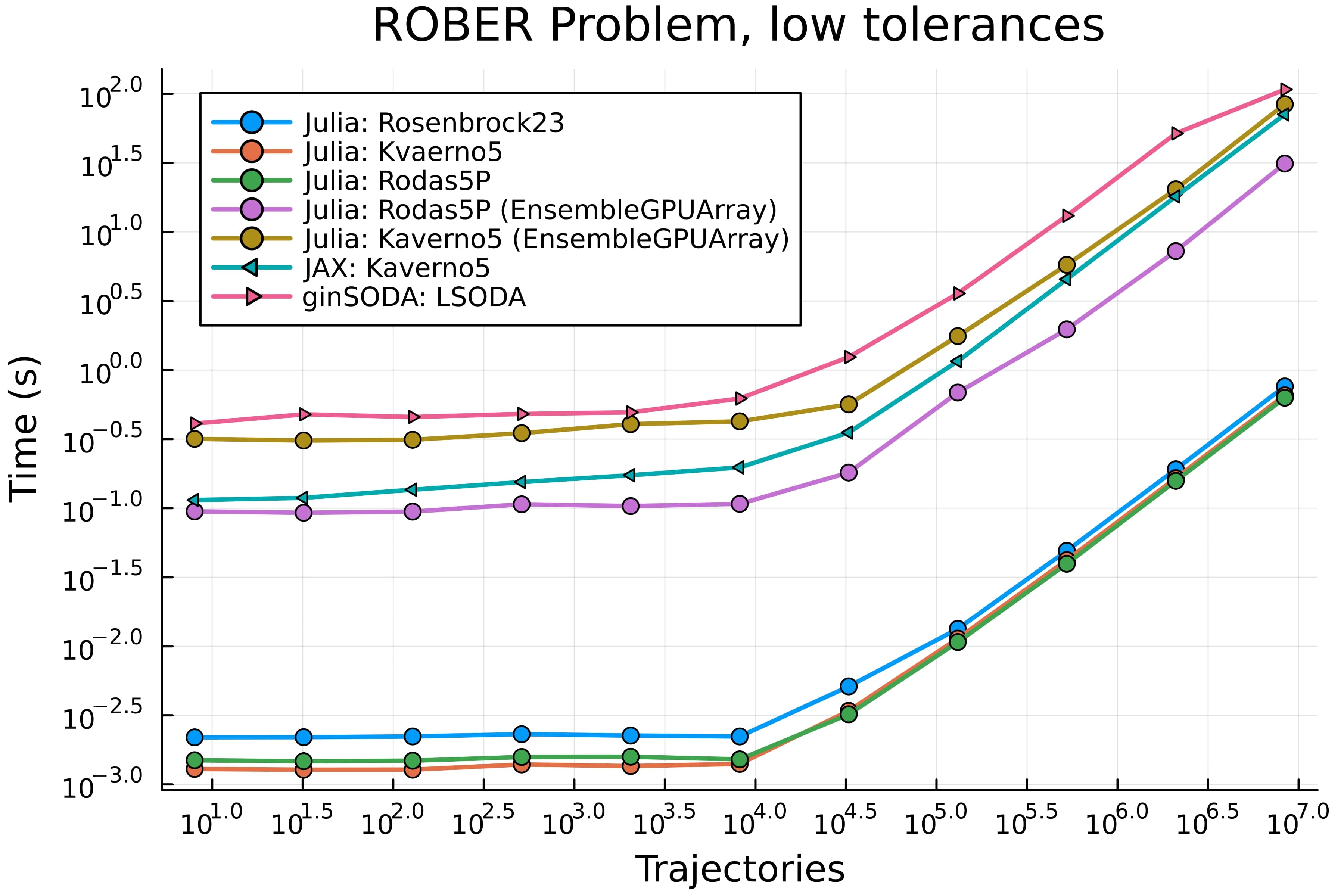}
  \caption{A comparison of the time for a stiff ODE solve with other programs with adaptive time-stepping. EnsembleGPUKernel is faster than JAX and EnsembleGPUArray by approximately 76--130$\times$.}
  \label{benchmark:rober_adaptive}
\end{figure}

Similarly, we benchmark the solvers with stiff ODEs. A usual test case for stiff ODEs is the Robertson equation \cite{robertson1976numerical}, which models a chemical reaction between three species. The rate of the reactions varies at drastically different time scales, which subsequently gives rise to the stiffness characteristic of the problem. We also additionally benchmark with an other massively parallel stiff ODE solver, ginSODA \cite{s2019ginsoda}, which uses Python as a front-end for their CUDA-C++ kernels. Figure \ref{benchmark:rober_adaptive} summarizes the benchmarks when compared with JAX, Julia's EnsembleGPUArray, and ginSODA.
The performance trend remains congruent with non-stiff ODE problems, indicating an average speed-up of 76--130$\times$ for JAX, 129--280$\times$ for Julia's EnsembleGPUArray, and 170--409$\times$ for ginSODA. According to the author's knowledge, the stiff ODE solvers are the first most extensive and the state-of-the-art massively parallel kernel-based solvers available for GPU acceleration.

\begin{table*}
    \caption{A summary of the range of slowdowns of the benchmarks in Figures \ref{benchmark:lorenz_unadaptive} and \ref{benchmark:lorenz_adaptive} \textit{(lower is better)}, with results compiled on a desktop GPU. The slowdowns are computed by varying the number of trajectories. The Julia-based solvers achieve the best acceleration on average.}
    \centering
    \begin{tabular}{|c|c|c|}
           \toprule
            \backslashbox{Software}{Time-stepping} & Fixed & Adaptive  \\
           \midrule
        DiffEqGPU.jl (Julia) & 1.0$\times$ & 1.0$\times$ \\ 
        MPGOS (C++) & 2.2--$5.3\times$ & 0.5--$2.3\times$ \\
        Diffrax (JAX) & 88.9--274.0$\times$ & 28.0--$124.0\times$ \\
        torchdiffeq (PyTorch) & 196.4--3617.7$\times$ &  --- \\
        \bottomrule
    \end{tabular}
    
    \label{tab:impl_comparison_rtx}
\end{table*}

\begin{table*}
    \caption{A summary of the range of slowdowns of the benchmarks in Figures \ref{benchmark:lorenz_unadaptive} and \ref{benchmark:lorenz_adaptive} \textit{(lower the better)}, with results compiled on a server GPU. The slowdowns are computed by varying the number of trajectories. The Julia-based solvers achieve the best acceleration on average.}
   \centering
    \begin{tabular}{|c|c|c|}
           \toprule
            \backslashbox{Software}{Time-stepping} & Fixed & Adaptive  \\
           \midrule
        DiffEqGPU.jl (Julia) & 1.0$\times$ & 1.0$\times$ \\ 
        MPGOS (C++) & 0.6--10.0$\times$ & 0.6--6.9$\times$ \\
        Diffrax (JAX) & 57.4--322.3$\times$ & 30.2--46.8$\times$ \\
        torchdiffeq (PyTorch) & 98.8--3693.9$\times$ &  --- \\
        \bottomrule
    \end{tabular}
    \label{tab:impl_comparison_v100}
\end{table*}


\begin{table*}
    \caption{A summary of the range of slowdowns of the benchmarks in Figure \ref{benchmark:rober_adaptive} \textit{(lower the better)}, with results compiled on a server GPU for stiff ODEs. The slowdowns are computed by varying the number of trajectories. The Julia-based solvers achieve the best acceleration on average.}
   \centering
    \begin{tabular}{|c|c|c}
           \toprule
            \backslashbox{Software}{Time-stepping} & Adaptive  \\
           \midrule
        EnsembleGPUKernel (Rodas5P) &  1.0$\times$ \\
        EnsembleGPUKernel (Kvaerno5) &  0.85--1.0$\times$ \\
        EnsembleGPUKernel (Rosenbrock23) &  1.2--1.6$\times$ \\
        EnsembleGPUArray (Kvaerno5) & 129.1--279.9$\times$ \\
        Diffrax (Kvaerno5) & 76.6--130.0$\times$ \\
        ginSODA (LSODA) & 170.4--409.4$\times$ \\
        \bottomrule
    \end{tabular}
    \label{tab:impl_comparison_stiff_v100}
\end{table*}

\subsection{Vendor agnosticism with performance: Comparison with several GPU platforms}

\begin{figure}[ht]
  \centering
  \includegraphics[width=0.7\linewidth]{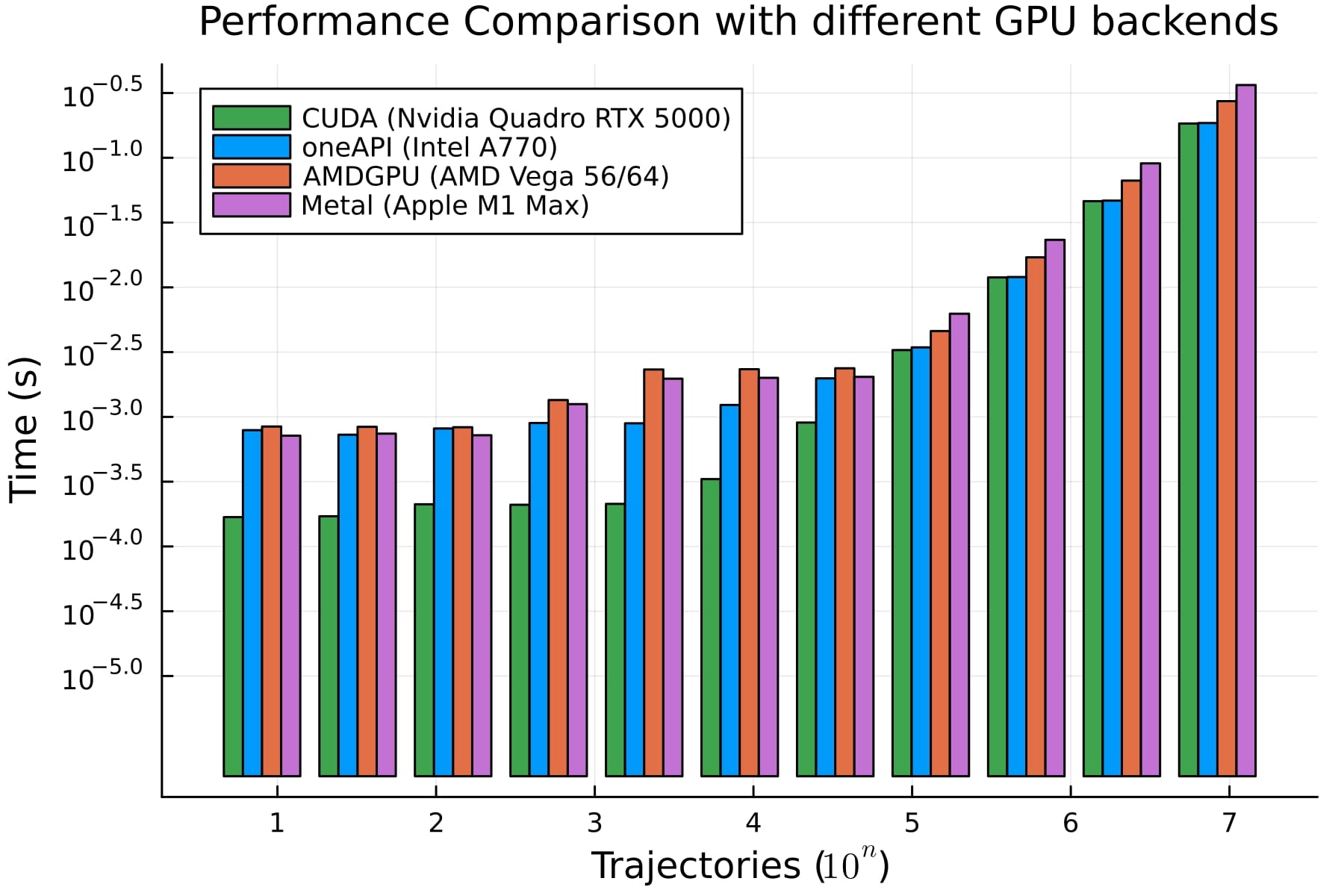}
  \caption{A comparison of the time for an ODE solve for fixed time-stepping, measured on different GPU platforms. The non-stiff ODE solver \texttt{GPUTsit5} is used here.
  We measure the time \textit{(lower the better)} versus the number of parallel solves. Here, the NVIDIA GPUs perform the best owing to the most-optimized library and matured ecosystem with JuliaGPU.}
  \label{benchmark:multi_gpu}
\end{figure}

With vendor-agnostic GPU kernel generation, researchers can choose major GPU backends with ease. Our benchmarks in Figure \ref{benchmark:multi_gpu} demonstrate that overhead is minimal in our ODE solvers and users can expect performance one-to-one with mentioned Floating Point Operations per Second (FLOPS) in GPUs mentioned in the Section \ref{section:setup}. To elude other performance impacts such as thread divergence and equitable selection of the dimension of the ODE, we simulate the Lorenz problem with fixed time-stepping. We run the benchmarks on major vendors: NVIDIA, AMD, Intel, and Apple. The peak flops are listed in Section \ref{section:setup}. To our knowledge, this is the first showcase of GPU-parallel software for solving DEs that supports more than NVIDIA GPUs.

\begin{figure}[ht]
  \centering
  \includegraphics[width=0.7\linewidth]{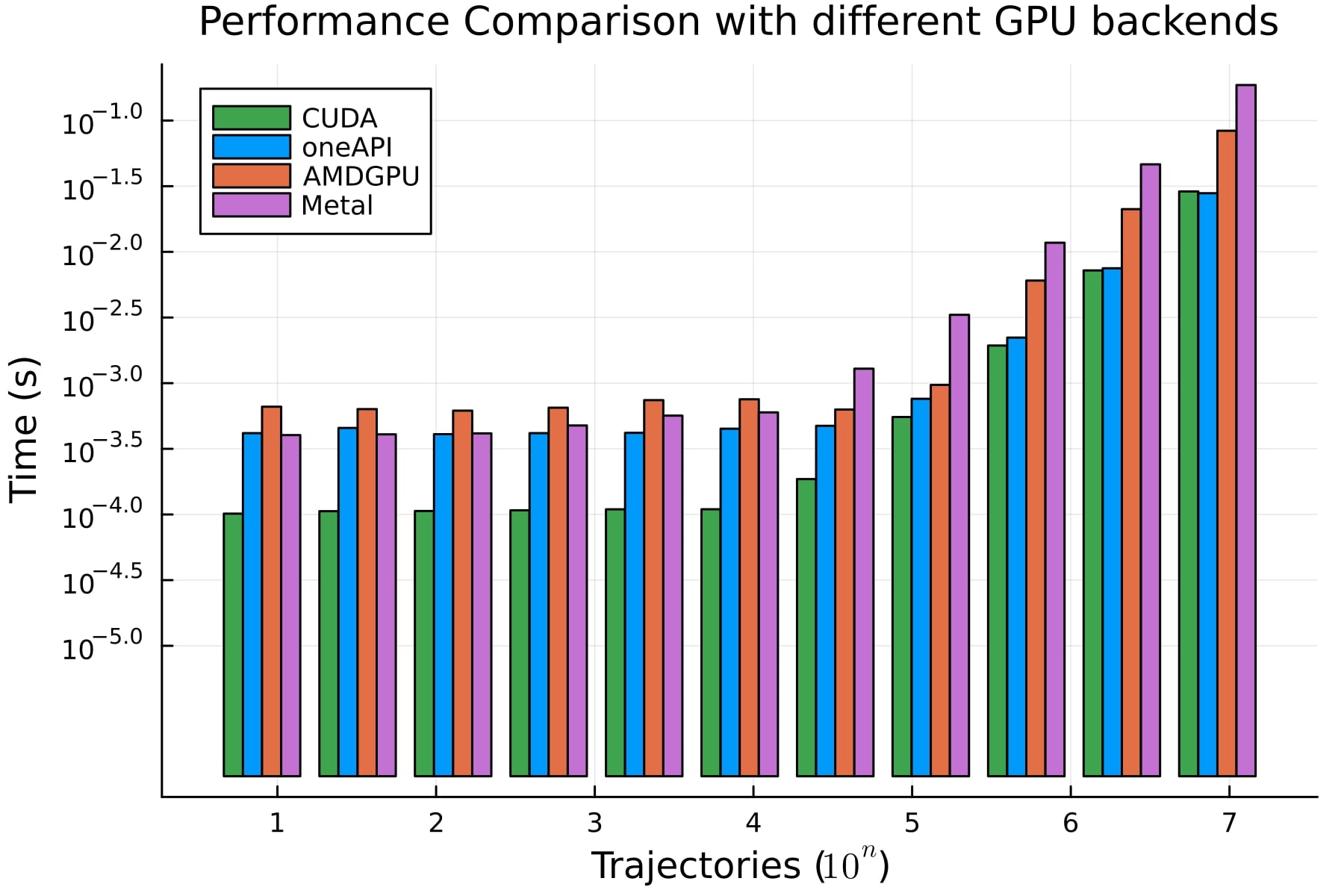}
  \caption{A comparison of of the time for an ODE solve for  adaptive time-stepping, measured on different GPU platforms. The stiff ODE solver \texttt{GPURosenbrock23} is used here. We measure the time \textit{(lower the better)} versus the number of parallel solves. Here, the NVIDIA GPUs perform the best owing to the most-optimized library and matured ecosystem with JuliaGPU.}
  \label{benchmark:multi_gpu_adaptive}
\end{figure}

\subsection{Event handling and automatic differentiation}

\begin{figure}[ht]
  \centering
  \includegraphics[width=0.7\linewidth]{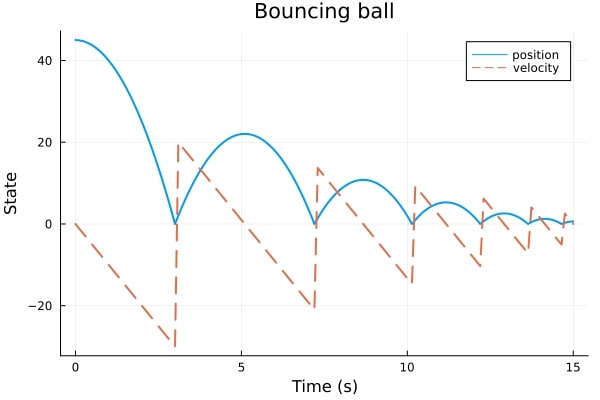}
  \caption{A simulation of bouncing ball problem on GPU. The blue trajectory is the displacement and the red trajectory is the velocity, across time. This demonstrates the ability to inject code within ODEs via callbacks.}
  \label{fig:bouncing_ball}
\end{figure}

Software for simulating dynamical systems allows for injecting discontinuous events and termination of integration with a specified criterion causing discontinuities within the integration. Event handling in differential equations is used to produce non-differentiable points in continuous dynamical systems. Mathematically, an event within an ODE can be specified as a tuple of functions, $g, h$ ("condition" and "affect"), where satisfying the "condition" $g(u,p,t) = 0$ triggers the "affect" $h(u,p,t)$, changing $u,t$ or terminating the integration. As a demonstration of event capabilities with \texttt{EnsembleGPUArray} and \texttt{EnsembleGPUKernel}, we demonstrate the famous surface-ball collision (bouncing ball) dynamics simulated on GPUs in Figure \ref{fig:bouncing_ball}.

Additionally, the GPU kernels are automatic differentiation (AD) compatible, both with forward with reverse mode, allowing for GPU-parallel forward and adjoint sensitivity analysis. Tutorials in the library demonstrate the usage of AD for parameter estimation with minibatching.\footnote{\href{https://docs.sciml.ai/SciMLSensitivity/stable/tutorials/data_parallel/\#Minibatching-Across-GPUs-with-DiffEqGPU}{https://docs.sciml.ai/SciMLSensitivity/stable/tutorials/data\_parallel/\#Minibatching-Across-GPUs-with-DiffEqGPU}}

\subsection{Texture memory interpolation}

As the mathematical model used to simulate a dynamical system is oftentimes a low-fidelity approximation of the true physical phenomena, practitioners often account for non-modeled behavior via lookup tables and interpolation of datasets. These datasets may be derived from real-world experimentation and/or higher-fidelity simulation. For cases where the interpolant is a function of the system state, interpolation is required at each time step for each system in the ensemble. A simple example of this is the above bouncing ball problem, extended to include drag forces imparted on the ball via the interpolation of a spatially varying wind field. Furthermore, a user may need to interpolate digital terrain elevation data for ground collision event handling, in which the textured memory defined in Listing \ref{Texture memory Example} can be used. 

\begin{lstlisting}[caption={An outline of using texture memory with DifferentialEquations.jl},captionpos=b,label = {Texture memory Example}]
    // Building textured memory of the dataset
    texture = CuTexture(CuTextureArray(dataset))
    // Passing the textured memory as parameter of the problem
    prob = ODEProblem(f_rhs, u0, tspan, (p, texture))
    // Performing the solve, with 0 CPU offloading.
    sol = solve(prob, alg, EnsembleGPUKernel(0), ...)
\end{lstlisting}

In addition to exploiting parallelism for time-stepping as described above, GPU texture memory can be leveraged when interpolation is required with multiple benefits. In particular, for NVIDIA GPUs, texture memory provides interpolation, nearest-neighbor search, and automatic boundary handling for the cost of a single memory read. Benchmarking with texture memory results in $2\times$ faster simulation, when we replace a compute-bound operation such as interpolation on GPU with texture-memory.

Furthermore, texture memory is advantageous for situations where the memory access pattern is not known \textit{a priori}, which is often the case for state-dependent interpolation. Texture memory, however, requires uniformly spaced data.

\subsection{Accelerating stochastic processes with GPUs}

The expectation of SDE solutions is a key metric in many model analyses due to the randomness of the simulation process. Such a calculation is generally done through Monte Carlo estimation via generating many trajectories of SDE solution, typically requiring  tens of thousands of trajectories for suitable convergence due to the $\mathcal{O}(\sqrt{N})$ convergence rate to the expectation. The generation of these trajectories is independent of each other, thus fitting the form of ensemble parallelism with the same initial condition and parameter values, with the only difference being the seed supplied to the Pseudorandom Number Generator (PRNG).

\subsubsection{Asset Price Model in Quantitative Finance} \label{sssection: linear sde}

\begin{figure}[ht]
  \centering
  \includegraphics[width=0.7\linewidth]{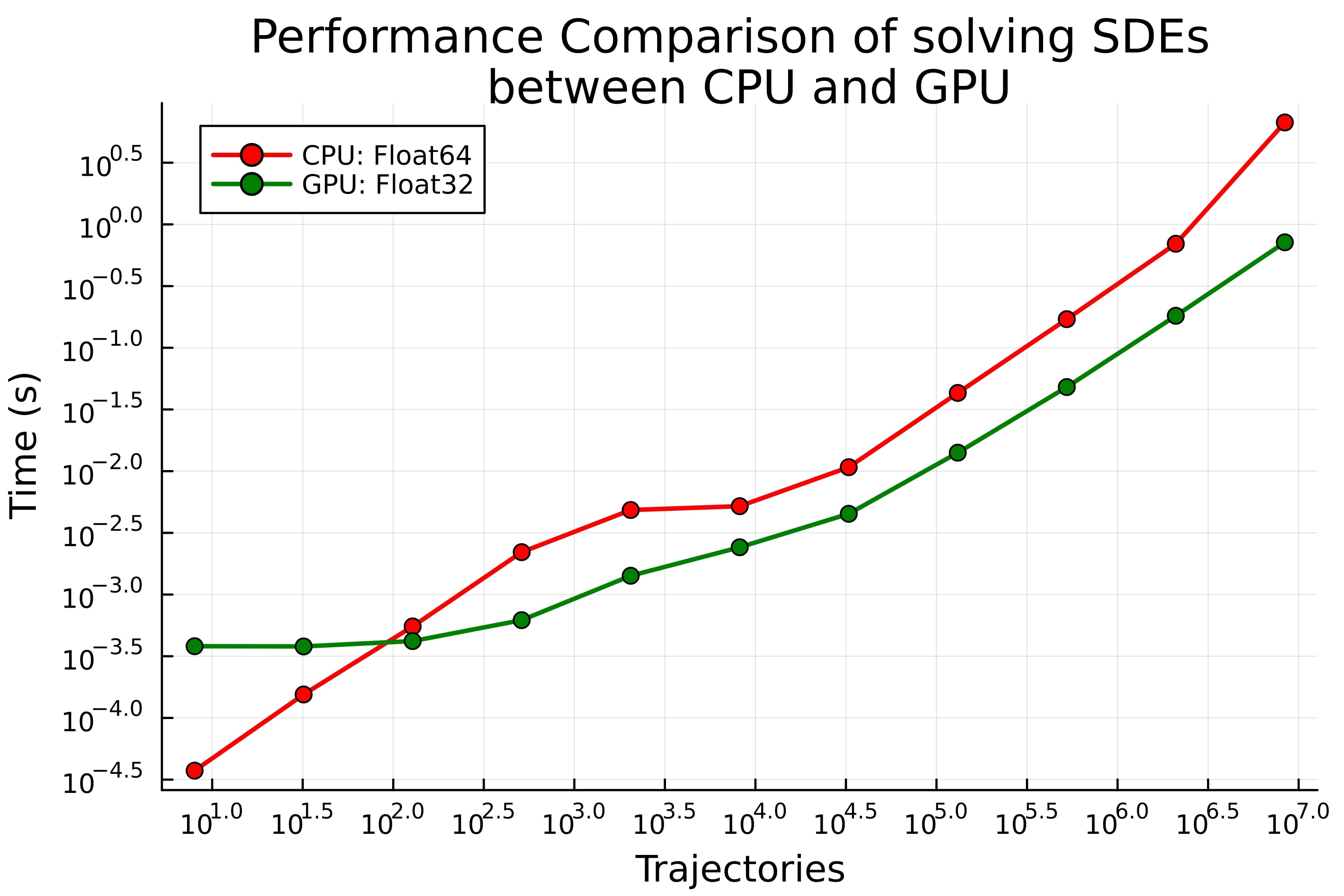}
  \caption{The parameter parallel simulation time for the linear SDE defined in Section \ref{sssection: linear sde} \textit{(lower the better)}. The GPU parallelism supercedes CPU parallelism at about 1000 trajectories.}
  \label{benchmark:gpu_sde_linear}
\end{figure}

As a rudimentary example, we examine the ensemble simulation of a linear SDE, popularly known as Geometric Brownian Motion (GBM), which is the common Black-Scholes model used in asset pricing in quantitative finance \cite{black1973pricing, merton1973theory}. The benchmarks in Figure \ref{benchmark:gpu_sde_linear} demonstrate that simulations on GPU are faster than CPU on average by $8\times$. GPU parallelism dominates CPU parallelism over approximately $1000$ trajectories. We note that the decreased performance difference from the ODE case is likely due to the less optimized implementation of the kernel PRNG implementations.

\subsubsection{Stochastic Chemical Reaction Networks}
\begin{figure}[ht]
  \centering
  \includegraphics[width=0.7\linewidth]{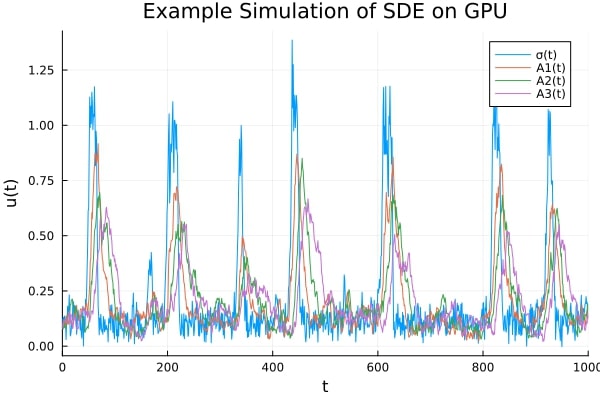}
  \caption{An example simulation plot of the system vs.\ time. The model is written with Catalyst.jl and automatically works with GPU solvers, showcasing the ability to simulate complex models seamlessly on GPU.}
  \label{fig:gpu_sde_MTK}
\end{figure}

\begin{figure}[ht]
  \centering
  \includegraphics[width=0.7\linewidth]{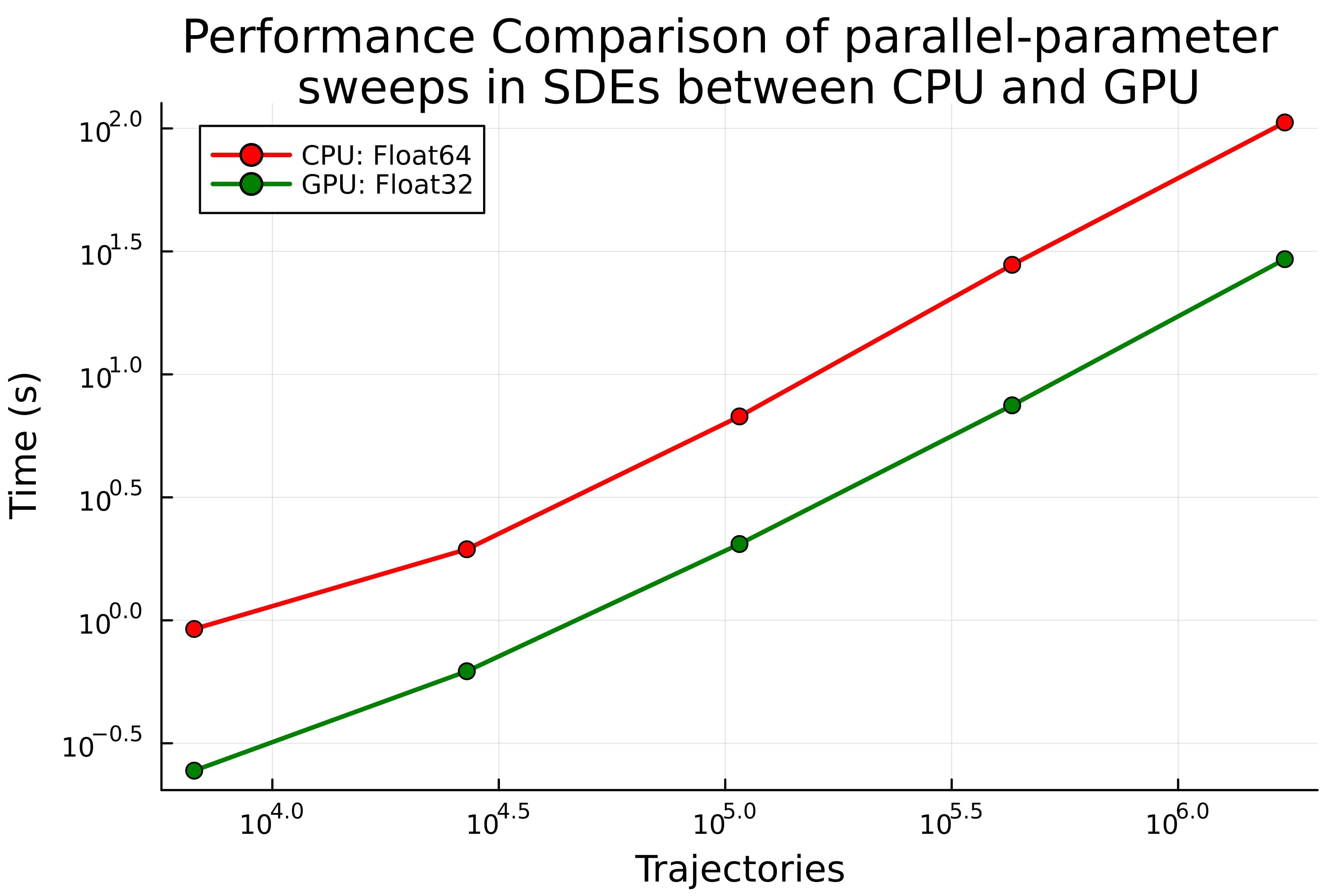}
  \caption{The parameter parallel simulation time \textit{(lower the better)} of the SDE simulation of Figure \ref{fig:gpu_sde_MTK}. Overall, the comparison showcases the scalability of speedups of using GPUs instead of CPUs, having suitable gains for trajectories as small as $1000$.}
  \label{benchmark:gpu_sde_MTK}
\end{figure}

Additionally, we benchmark the SDE solvers over a real case study of Chemical Reaction Networks (CRN) generated when microorganisms such as bacteria respond to stimuli, which causes a change in its gene expression through sigma factors \cite{torkelthesis}. These biological processes are inherently noisy in nature, and it is simulated by transforming the CRN to an SDE via the Chemical Langevin Equation (CLE) \cite{gillespie2000chemical}. The non-dimensionalized model has notably 4 states, 8 Wiener noise variables, and 6 parameters, making it suitable for our case study as our GPU DE solvers are suited for problems with low-dimensional states. The simulation of the process on the GPU in shown in Figure \ref{fig:gpu_sde_MTK}. The benchmark investigates the generation of trajectories of solutions for different parameters akin to parameter sweeps, which are widely used in parameter estimation and uncertainty quantification. Each of the parameters is uniformly sampled and the set of the Cartesian products of parameters is simulated, generating approximately $>$1,000,000 unique trajectories. The benchmarks shown in Figure \ref{benchmark:gpu_sde_MTK} quantitatively shows that our GPU implementation for SDEs is $4.5\times$ faster than multithreading over a CPU, averaged over different trajectories.

\section{Discussion}

We have demonstrated that many programs written for standard CPU usage of DifferentialEquations.jl can be retargeted to GPUs via DiffEqGPU.jl and achieve state-of-the-art (SOA) performance without requiring changes to user code. This solution democratizes the SOA by not requiring scientists and engineers to learn CUDA C++ in order to achieve top performance. One key result of the paper is that we demonstrate that all approaches which used array abstraction GPU parallelism, PyTorch, JAX, and \texttt{EnsembleGPUArray}, achieve similar performance and are orders of magnitude less efficient than the kernel generation approach of \texttt{EnsembleGPUKernel} and MPGOS. This suggests that the performance difference is due to a limitation of the array abstraction parallelization formulation, and demonstrates a concrete application where kernel generation is required for achieving SOA. It has previously been noted that ``machine learning systems are stuck in a rut'' where many deep learning architectures are designed to only use the kernels included by current machine learning libraries (PyTorch and JAX) \cite{10.1145/3317550.3321441}. Our results further this thesis by demonstrating that orders of magnitude performance improvements can only be achieved by leaving the constrained array-based DSL of pre-defined kernels imposed by such deep learning frameworks.

Despite being the performant alternative of \texttt{EnsembleGPUArray}, there are opportunities for improvements with \texttt{EnsembleGPUKernel}. For example, while using stack-allocated arrays provides a workaround to using arrays inside GPU kernels, they are not suitable for higher-dimensional problems due to the limited memory of static allocations. The model might compile, but there might not be any realizable speedups. Flexibility in terms of supporting mutation within the ODE function can be extended as well. This could be achieved by using mutable static arrays, which require special tricks to compile them with the GPU kernels. The user is also limited in terms of using features such as broadcast and calls to BLAS. Support for other differential equation problems such as differential algebraic equations (DAEs) and stiff SDEs remains an area of contribution. Experimental support exists for event handling, however, some callbacks can generate GPU-incompatible code due to limitations in the Julia compiler. Improvements to the compiler's escape analysis and effects modeling are currently being implemented, and are expected to resolve this issue.

\section*{Acknowledgment}
We thank Mosè Giordano for developing an extension of our work to allow for using DiffEqGPU.jl on Graphcore Intelligence Processing Unit (IPU) architectures. This work is omitted from our manuscript as it is a different development effort and has an alternative front-end, making it difficult to compare to the GPU architectures. However, his work allows for using the same DiffEqGPU.jl solver kernels recompiled to IPU targets, demonstrating similar good performance. We additionally thank Torkel Loman for his help with stochastic differential equation models. 

The authors acknowledge the MIT SuperCloud and Lincoln Laboratory Supercomputing Center for providing HPC resources that have contributed to the research results reported within this paper. This material is based upon work supported by the National Science Foundation under grant no. OAC-1835443, grant no. SII-2029670, grant no. ECCS-2029670, grant no. OAC-2103804, and grant no. PHY-2021825. The information, data, or work presented herein was funded in part by the Advanced Research Projects Agency-Energy (ARPA-E), U.S. Department of Energy, under Award Number DE-AR0001211 and DE-AR0001222. We also gratefully acknowledge the U.S. Agency for International Development through Penn State for grant no. S002283-USAID. The views and opinions of authors expressed herein do not necessarily state or reflect those of the United States Government or any agency thereof. This material was supported by The Research Council of Norway and Equinor ASA through the Research Council project ``308817 - Digital wells for optimal production and drainage.'' The research was sponsored by the U.S. Air Force Research Laboratory and Air Force Office of Scientific Research Lab Task \#21RQCOR083 in addition to the United States Air Force Artificial Intelligence Accelerator accomplished under Cooperative Agreement Number FA8750-19-2-1000. The views and conclusions contained in this document are those of the authors and should not be interpreted as representing the official policies, either expressed or implied, of the U.S. Air Force or the U.S. Government. The U.S. Government is authorized to reproduce and distribute reprints for Government purposes notwithstanding any copyright notation herein.

\bibliographystyle{elsarticle-num} 
\bibliography{refs}

\appendix

\section{Artifact Identification}

The manuscript aims to explore the parallel solving of the ensembles of ordinary differential equations (ODEs) and stochastic differential equations (SDEs) with GPUs. The methods automate the process by generating optimized GPU kernels without requiring any model changes by the user. The methods outperform the vectorized map approach adopted by JAX and PyTorch and even handwritten CUDA C++ kernels, becoming state-of-the-art for solving ensemble ODEs/SDEs. The solution is high-performant with vendor-agnosticism, working with NVIDIA, AMD, Intel, and Apple GPUs.

The paper evaluates the performance of the ODE and SDE solvers with benchmarking with CPU multithreading using Intel Xeon Gold 6248 CPU @ 2.50GHz with 16 enabled threads. The composability of the methods with Julia's SciML ecosystem is also demonstrated by simulating a stochastic process based on a chemical reaction network model, written in a Julia-based Domain Specific Language (DSL) Catalyst.jl. 

Further, we also benchmark against GPU-accelerated programs performed with ODE solvers. The GPUs used are Tesla V100 and RTX 500. Our case study used the software:
\begin{itemize} \label{itemize: programlist}
    \item EnsembleGPUArray (Julia's DifferentialEquations.jl)
    \item MPGOS (CUDA C++)
    \item JAX (Python)
    \item PyTorch (Python)
\end{itemize}

The \texttt{EnsembleGPUArray} based code is written in Julia and is a part of DiffEqGPU.jl, the MPGOS program is written in C++, and the benchmarking scripts for JAX and PyTorch are written in Python. A standard workflow to integrate the polyglot software for benchmarking is done via bash scripts, which invoke the benchmarking scripts and store the execution times in a ".txt" file. Finally, the plotting scripts for these benchmarks are written in Julia to visualize the numbers obtained through benchmarking. 

The description of the models used in benchmarking is provided in the next section.

\subsection{Models: Ordinary Differential Equations}

\subsubsection{Lorenz Problem} \label{lorenz_example} The first test problem is the Lorenz attractor:
\begin{align} \label{eq:lorenz}
&\frac{dy_1}{dt} = \sigma(y_2 - y_1),
\\
&\frac{dy_2}{dt} = \rho y_1 - y_2 - y_1y_3,
\\
&\frac{dy_3}{dt} = y_1y_2 - \gamma y_3,
\end{align}
which contains three parameters: $\sigma  = 10$, $\gamma = {8}/{3}$, and $\rho = 21$. The integration is performed for $t \in [0,1]$ with time-step $dt = 0.001$, essentially generating 1000 fixed time-steps. The initial conditions are $y = [1,0,0]$. 

\subsubsection{Bouncing-ball problem} \label{bouncing_ball_example} The problem simulated in the callbacks example is the bouncing ball problem which has the kinetics:
\begin{align}
&\frac{dx}{dt} = v,
\\
&\frac{dv}{dt} = -g,
\end{align}

\begin{lstlisting}[caption={Implementation of callbacks},captionpos=b,label = {lis:callback_example}]
    function affect!(integrator)
        # Flips the sign of velocity 
        # by coefficient of restitution
        integrator.u[2] = -integrator.e*integrator.u[2]
    end
    
    function condition(u, t, integrator)
        # returns true when the distance 
        # to the ground is evaluated as zero
        u[1] # == 0
    end
\end{lstlisting}

with $g = 9.8$ m/s$^2$ and $e$ is the coefficient of restitution, which varies across the simulation. The problem is simulated for time interval $[0,15]$. Completing the problem requires specification of how the object behaves when the ball hits the ground, which is done via event handling as described in Listing \ref{lis:callback_example}.

\subsubsection{Robertson Equation}

\label{rober_example} The test problem used for stiff ODEs is the Robertson Equation:
\begin{align} \label{eq:robertson}
&\frac{dy_1}{dt} = -0.04y_1 + 10^4y_2y_3,
\\
&\frac{dy_2}{dt} = 0.04y_1 - 10^4y_2y_3 - 3\mbox{$\times$}10^7 y_2^2,
\\
&\frac{dy_3}{dt} = 3\mbox{$\times$}10^7 y_2^2.
\end{align}

The integration is performed for $t \in [0.0,10^5]$ with the initial time-step of $h = 0.0001$. The initial conditions are $y = [1,0,0]$.

\subsection{Models: Stochastic Differential Equations}
\subsubsection{Linear SDE} \label{lin_sde_example} The asset-price model is simply expressed as
\begin{equation}
    dX_t = rX_tdt +  VX_tdW_t
\end{equation}
where $X_t$ is the state to be simulated with initial condition $X_{t_0} = [0.1,0.1,0.1]$, $r = 1.5$ is the risk-free rate of interest, and $V = 0.01$ is the volatility.

\subsubsection{Biological Chemical Reaction Network (CRN) simulating generalized stress response model of bacterial sFigma factors}
The biological CRN models the regulation of sigma factors via sigma factor circuits, which critically regulate gene expression during the bacterial stress response. These phenomena are stochastic in nature. The resulting process can be expressed as the SDE:
\begin{align}
d[\sigma] &= \left(\nu_0 + \frac{(S[\sigma])^n}{(S[\sigma])^n + (D[A3])^n + 1} - [\sigma]\right)dt + \eta\sqrt{\nu_0 + \frac{(S[\sigma])^n}{(S[\sigma])^n + (D[A3])^n + 1}}dW_1 - \eta\sqrt{[\sigma]}dW_2,
\\
d[A_1] &= \left(\frac{[\sigma]}{\tau} - \frac{[A_1]}{\tau}\right)dt 
       + \eta\sqrt{\frac{[\sigma]}{\tau}}dW_3 - \eta\sqrt{\frac{[A_1]}{\tau}}dW_4 ,
\\
d[A_2] &= \left(\frac{[A_1]}{\tau} - \frac{[A_2]}{\tau}\right)dt 
       + \eta\sqrt{\frac{[A_1]}{\tau}}dW_5 - \eta\sqrt{\frac{[A_2]}{\tau}}dW_6 ,
\\
d[A_3] &= \left(\frac{[A_2]}{\tau} - \frac{[A_3]}{\tau}\right) dt
       + \eta\sqrt{\frac{[A_2]}{\tau}}dW_7 - \eta\sqrt{\frac{[A_3]}{\tau}}dW_8,
\end{align}

\begin{table*}
    \caption{Range of parameters used in parameter-parallel SDE solutions}
    \label{tab:crn_sde_parameters}
    \centering
    \begin{tabular}{c|c}
    \toprule
    Parameter & Range \\
    \midrule
      $S$   & $0.1 \le S \le 100.0 $  \\
      $D$   & $0.1 \le D \le 100.0 $  \\
      $\tau$   & $0.1 \le \tau \le 100.0 $  \\
      $\nu_0$   & $0.01 \le S \le 0.2 $  \\
      $n$   & $2 \le S \le 4 $  \\
      $\eta$   & $0.001 \le S \le 0.1 $  \\
    \bottomrule
    \end{tabular}
\end{table*}

which is generated through the Chemical Langevin Equation (CRE). The parameters are the set 
$(S,D,\tau, \\ v_0,n, \eta)$. The range of parameters is specified in Table \ref{tab:crn_sde_parameters}. The simulation is performed for the time-span $[0,1000]$ with $dt = 0.1$ and initial condition $[[\sigma], [A_1], [A_2], [A_3]]$ being $[\nu_0,\nu_0,\nu_0,\nu_0]$.


The benchmarking measures the execution time of the methods in seconds. The timings are compared against the number of parallel solves (trajectories). Apart from benchmarks, we also present compatibility of the methods with standard scientific computing methodologies such as event-handling, automatic differentiation, HPC cluster scaling with MPI, and incorporation of the use of scientific datasets via textured memory. The methods are free to use via the open-source library, DiffEqGPU.jl, hosted on GitHub. The specific examples of the case studies and benchmarks are available on the complete benchmark suite:
\\\href{https://github.com/utkarsh530/GPUODEBenchmarks}{https://github.com/utkarsh530/GPUODEBenchmarks}. 

Together, these evaluations allow the paper to convey the SOA performance of the methods, scalability with distributed and multiple GPU vendors, reusability by retargeting existing CPU-targeted user's code to GPUs, and composability with other tools of the high-level programming language.

\section{Auxiliary model testing and benchmarks}

This section addresses the generalizability of the results of the GPU solvers with different differential equation models. 

\subsection{Model Description}

\subsubsection{OREGO} \label{section: orego}

The first test problem is the Oregonator (Orego) \cite{zhabotinsky2007belousov, field1974oscillations}.
\begin{align}
{}&\frac{dy_1}{dt} = -k_1(y_2 + y_1(1-k_2y_1 - y_2)),
\\
{}&\frac{dy_2}{dt} = \frac{y_3 - (1 + y_1)y_2}{k_1},
\\
{}&\frac{dy_3}{dt} = k_3(y_1 - y_3).
\end{align}
The initial conditions are $y = [1,2,3]$ and $k = (77.27,8.375$$\times$ $10^{-6},0.161)$. The time span for integration is $t \in [0,30]$\,s.

\subsubsection{HIRES} \label{section: hires}

The second test problem is HIRES \cite{schafer1975new}:
%
%
\begin{align}
\frac{dy_1}{dt} &= -1.71y_1 + 0.43y_2 + 8.32y_3 + 0.0007,
\\
\frac{dy_2}{dt} &= 1.71y_1 - 8.75y_2,
\\
\frac{dy_3}{dt} &= -10.03y_3 + 0.43y_4 + 0.035y_5,
\\ 
\frac{dy_4}{dt} &= 8.32y_2 + 1.71y_3 - 1.12y_4, 
\\ 
\frac{dy_5}{dt} &= -1.745y_5 + 0.43y_6 + 0.43y_7, 
\\ 
\frac{dy_6}{dt} &=  -280.0y_6y_8 + 0.69y_4
                + 1.71y_5 - 0.43y_6 + 0.69y_7, 
\\
\frac{dy_7}{dt} &= 280.0y_6y_8 - 1.81y_7,
\\
\frac{dy_8}{dt} &= -280.0y_6y_8 + 1.81y_7.
\end{align}
The initial conditions are:
\begin{equation}
 y = [1.0,0.0,0.0,0.0,0.0,0.0,0.0,0.0057].
\end{equation}
The time span for integration is $t \in [0.0,321.8122]$\,s.

\subsubsection{POLLU} \label{section: pollu}
The third test problem is the pollution (POLLU) problem \cite{verwer1994gauss}:
%
%
\begin{align}
\frac{dy_1}{dt} &= -k_1y_1 - k_{10}y_{11}y_1 - k_{14}y_1y_6 - k_{23}y_1y_4
                - k_{24}y_{19}y_1 + k_2y_2y_4 + k_3y_5y_2 + k_9y_{11}y_2 \\
                & + k_{11}y_{13} + k_{12}y_{10}y_2 + k_{22}y_{19} + k_{25}y_{20},
\\
\frac{dy_2}{dt} &= -k_2y_2y_4 - k_3y_5y_2 - k_9y_{11}y_2 - k_{12}y_{10}y_{2} 
                + k_1y_1 + k_{21}y_{19},
\\
\frac{dy_3}{dt} &= -k_{15}y_3 + k_1y_1 + k_{17}y_4 + k_{19}y_{16} + k_{22}y_{19},
\\ 
\frac{dy_4}{dt} &= -k_2y_2y_4 - k_{16}y_4 - k_{17}y_4 - k_{23}y_1y_4 + k_{15}y_3, 
\\ 
\frac{dy_5}{dt} &= -k_3y_5y_2 + 2k_4y_7 + k_6y_7y_6 + k_7u_9 \\
                & + k_{13}y_{14} + k_{20}y_{17}y_6, 
\\ 
\frac{dy_6}{dt} &= -k_6y_7y_6 - k_8y_9y_6 - k_{14}y_1y_6 - k_{20}y_{17}y_6
                + k_3y_5y_2 + 2k_{18}u_{16}, 
\\
\frac{dy_7}{dt} &= -k_4y_7 - k_5y_7 - k_6y_7y_6 + k_{13}y_{14},
\\
\frac{dy_8}{dt} &= k_4y_7 + k_5y_7 + k_6y_7y_6 + k_7y_9,
\\
\frac{dy_9}{dt} &= -k_7y_9 - k_8y_9y_6,
\\
\frac{dy_{10}}{dt} &= -k_{12}y_{10}y_2 + k_7y_9 + k_9y_{11}y_2,
\\
\frac{dy_{11}}{dt} &= -k_9y_{11}y_2 - k_{10}y_{11}y_1 + k_8y_9y_6 + k_{11}y_{13},
\\ 
\frac{dy_{12}}{dt} &= k_9y_{11}y_2, 
\\ 
\frac{dy_{13}}{dt} &= -k_{11}y_{13} + k_{10}y_{11}y_1, 
\\ 
\frac{dy_{14}}{dt} &=  -k_{13}y_{14} + k_{12}y_{10}y_2, 
\\
\frac{dy_{15}}{dt} &= k_{14}y_1y_6,
\\
\frac{dy_{16}}{dt} &= -k_{18}y_{16} - k_{19}y_{16} + k_{16}y_4.
\\
\frac{dy_{17}}{dt} &= -k_{20}y_{17}y_6, 
\\ 
\frac{dy_{18}}{dt} &=  k_{20}y_{17}y_6, 
\\
\frac{dy_{19}}{dt} &= -k_{21}y_{19} - k_{22}y_{19} - k_{24}y_{19}y_1 + k_{23}y_1y_4 + k_{25}y_{20},
\\
\frac{dy_{20}}{dt} &= -k_{25}y_{20} + k_{24}y_{19}y_1.
\end{align}
\begin{align}
k = &[0.35, 26.6, 12300.0, 0.00086, 0.00082, 15000.0, 0.00013, 24000.0, 16500.0, 9000.0, 0.022, 12000.0, \\& 1.88, 16300.0, 4.8e6, 0.00035, 0.0175, 1.0\mbox{$\times$}10^8, 4.44\mbox{$\times$}10^{11}, 1240.0, 2.1, 5.78, 0.0474, 1780.0, 3.12]. 
\end{align}
\begin{align}
y =&[0.0,0.2,0.0,0.04,0.0,0.0,0.1,0.3,0.017,0.0,0.0,0.0,0.0,0.0,0.0,0.0,0.007,0.0,0.0,0.0].
\end{align}
The time span for integration is $t \in [0.0,60.0]$\,s.

\subsection{Preserving performance with different ODE models}

\begin{figure}[ht]
  \centering
  \includegraphics[width=0.7\linewidth]{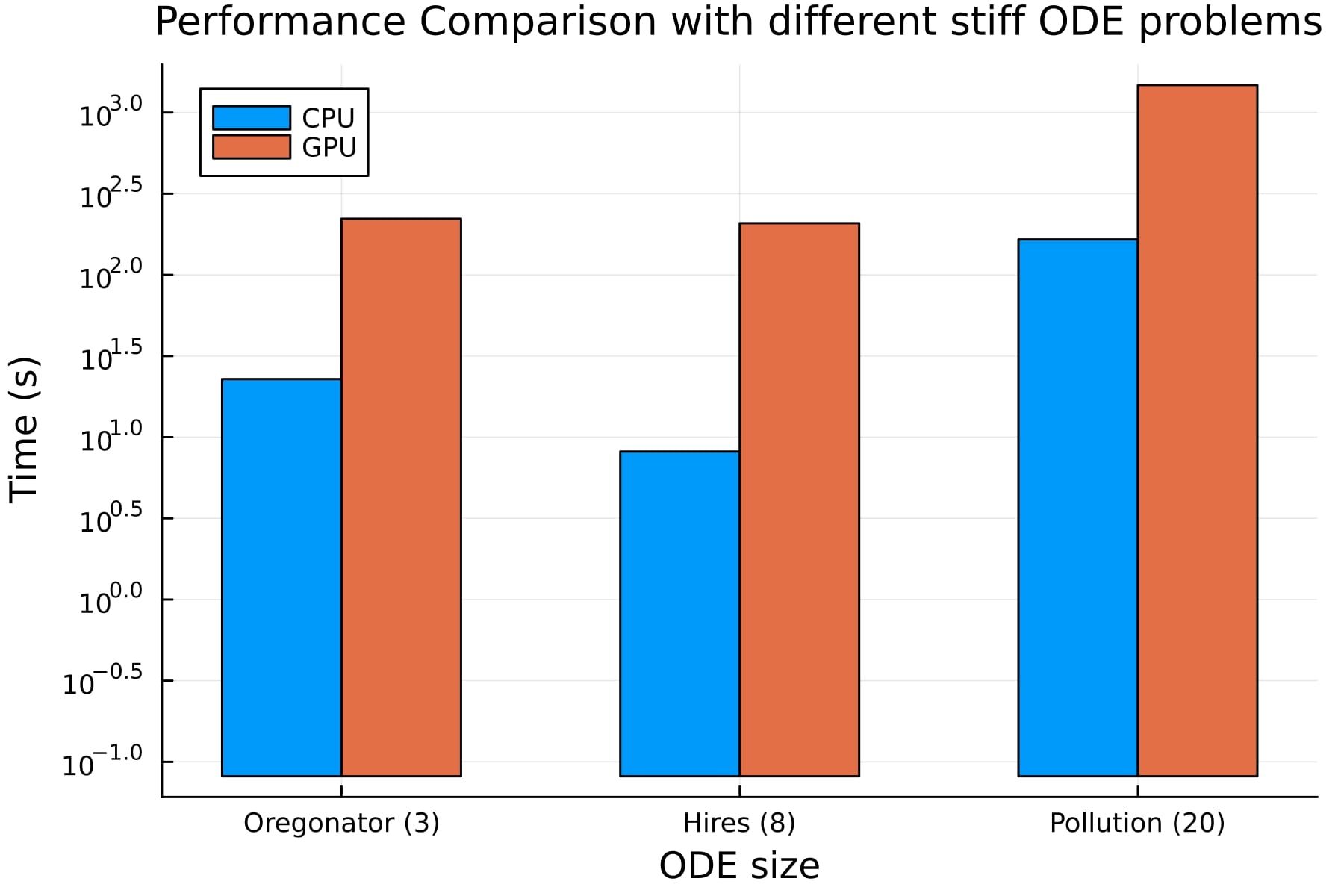}
  \caption{A comparison of the time for an ODE solve \textit{(lower the better)} with different stiff ODE problems having different sizes. The average speedup over CPU parallelism at 8192 trajectories is approximately 10$\times$.}
  \label{benchmark:scaling_stiff_ode}
\end{figure}

In this section, we benchmark the ODE solvers on different stiff problems, each having varying dynamics and sizes of the model. We benchmark our solvers with three problems \ref{section: hires}, \ref{section: orego}, \ref{section: pollu}. To allow GPU parallelism to sufficiently parallelize computations so as to overcome overheads such as kernel launch times, 8192 trajectories of the solution are computed. Figure \ref{benchmark:scaling_stiff_ode} showcases the average speedup of approximately 10$\times$ on all problems in comparison to the CPU parallelism. The benchmark demonstrates the scalability of the solvers with different ODE problems.

\subsection{Benchmarking different GPU solvers with AMD GPUs}

\begin{figure}[ht]
  \centering
  \includegraphics[width=0.7\linewidth]{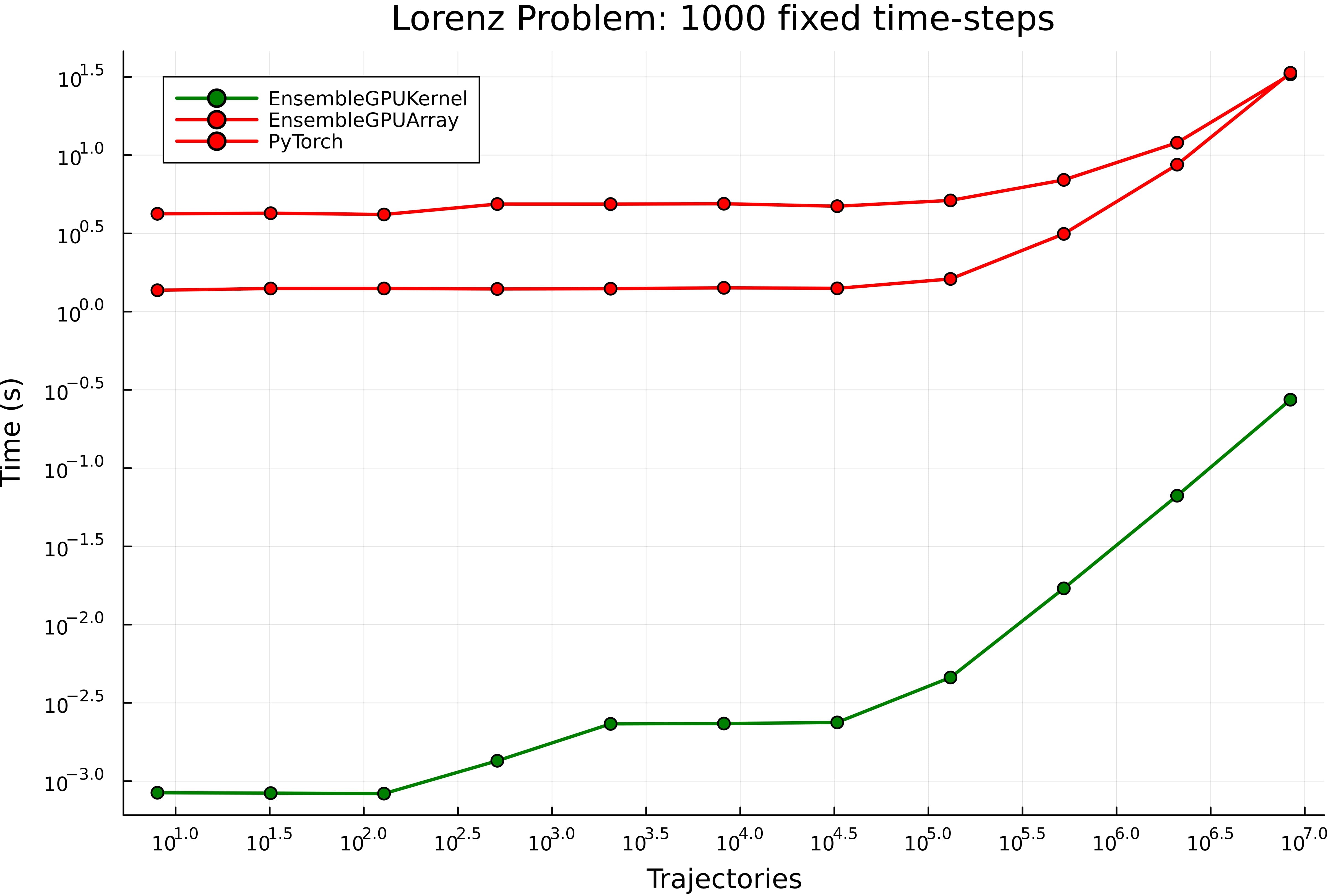}
  \caption{A comparison of the time for an ODE solve for other programs with fixed time-stepping, using AMD GPUs. Here, the \texttt{EnsembleGPUKernel} is at least 100$\times$ faster than \texttt{EnsembleGPUArray} and \texttt{PyTorch}.}
  \label{benchmark:lorenz_unadaptive_amd}
\end{figure}

Section \ref{section: comp_gpu_programs} covers benchmarking with different software with NVIDIA GPUs. Here, we present our results for AMD GPUs, which are currently fully supported by PyTorch and Julia's DiffEqGPU.jl. Figure \ref{benchmark:lorenz_unadaptive_amd} shows the comparison of the Lorenz problem, achieving at least 100$\times$ speedup. This demonstrates that the performance is independent of the GPU vendor and does not employ any specialized techniques such as CUDA GEMM kernels, allowing scientists to employ a variety of GPUs for their computation.

\section{Reproducibility of Experiments}

The methods are written in Julia, and are part of the repository, \href{https://github.com/SciML/DiffEqGPU.jl}{\path{https://github.com/SciML/DiffEqGPU.jl}}. The benchmark suite also consists of the raw data, such as simulation times and plots mentioned in the paper. The supported OS for the benchmark suite is Linux.

\subsection{Installing Julia}

The first step is to install Julia. The user can download the binaries from the official JuliaLang website \href{https://julialang.org/downloads/}{\nolinkurl{https://julialang.org/downloads/}}. or alternatively use the convenience of a Julia version multiplexer, \href{https://github.com/JuliaLang/juliaup}{\nolinkurl{https://github.com/JuliaLang/juliaup}}. The recommended OS for installation is Linux. The recommended Julia installation version is v1.8. To use AMD GPUs, please install v1.9. The Julia installation should also be added to the user's path.

\subsection{Setting up DiffEqGPU.jl}

\subsubsection{Installing backends}

The user must install the GPU backend library for testing DiffEqGPU.jl-related code.
\begin{verbatim}
julia> using Pkg
julia> #Run either of them
julia> Pkg.add("CUDA") # NVIDIA GPUs
julia> Pkg.add("AMDGPU") #AMD GPUs
julia> Pkg.add("oneAPI") #Intel GPUs
julia> Pkg.add("Metal") #Apple M series GPUs
\end{verbatim}

\subsubsection{Testing DiffEqGPU.jl}
DiffEqGPU.jl is a test suite that regularly checks functionality by testing features such as multiple backend support, event handling, and automatic differentiation. To test the functionality, one can follow the below instructions. The user needs to specify the "backend" for example "CUDA" for NVIDIA, "AMDGPU" for AMD, "oneAPI" for Intel, and "Metal" for Apple GPUs. The estimated time of completion is 20 minutes.

\begin{verbatim}
$ julia --project=.
julia> using Pkg
julia> Pkg.instantiate()
julia> Pkg.precompile()
\end{verbatim}

Finally test the package by this command

\begin{verbatim}
$ backend="CUDA"
$ julia --project=. test_DiffEqGPU.jl $backend
\end{verbatim}

Additionally, the GitHub discussion \href{https://github.com/SciML/DiffEqGPU.jl/issues/224\#issuecomment-1453769679}{\nolinkurl{https://github.com/SciML/DiffEqGPU.jl/issues/224}} highlights the use of textured memory with ODE solvers, accelerated the code by $2\times$ over GPU.

\subsubsection{Continuous Integration and Development}

DiffEqGPU.jl is a fully featured library with regression testing, semver versioning, and version control. The tests are performed on cloud machines having a multitude of different GPUs,  \href{https://buildkite.com/julialang/diffeqgpu-dot-jl/builds/705}{\nolinkurl{https://buildkite.com/julialang/diffeqgpu-dot-jl/builds/705}}. These tests approximately complete in 30 minutes. The publicly visible testing framework serves as a testimonial of compatibility with multiple platforms and said features in the paper.

\subsection{Testing GPU-accelerated ODE Benchmarks with other programs}

\subsubsection{Benchmarking Julia (DiffEqGPU.jl) methods} \label{subsubsection: bench_julia}

CUDA.jl needs to be installed for benchmarking as it is the only backend compatible with the ODE solvers in JAX, PyTorch, and MPGOS. To do so, enter 
\begin{verbatim}
$ julia
julia> using Pkg
julia> Pkg.add("CUDA")
\end{verbatim}
in the Julia Terminal.

Clone the benchmark suite repository to start benchmarking;

\begin{verbatim}
$ git clone https://github.com/utkarsh530\
/GPUODEBenchmarks.git
\end{verbatim}

We will instantiate and pre-compile all the packages beforehand to avoid the wait times during benchmarking. The folder ./GPU\_ODE\_Julia contains all the related scripts for the GPU solvers,

\begin{verbatim}
$ cd ./GPUODEBenchmarks
$ julia --project=./GPU_ODE_Julia --threads=auto
julia> using Pkg
julia> Pkg.instantiate()
julia> Pkg.precompile()
julia> exit()
\end{verbatim}

It may take a few minutes to complete (< 10 minutes). After this, we can generate the timings of ODE solvers written in Julia. There is a script to benchmark ODE solvers for the different number of trajectories to demonstrate scalability and performance. The script invocation and timings can be generated through
\begin{verbatim}
$ bash ./run_benchmark.sh -l julia -d gpu -m ode
\end{verbatim}
which might take about 20 minutes to finish. The flag \texttt{-n N} can be used to specify the upper bound of the trajectories to benchmark. By default $N = 2^{24}$, where the simulation runs for $n \in 8 \le n < N$, with the multiples of $4$.

The data will be generated in the \texttt{data/Julia} directory, with two files for fixed and adaptive time-stepping simulations. The first column in the ".txt" file will be the number of trajectories, and the section column will contain the time in milliseconds.

Additionally, to benchmark ODE solvers for other backends:

\begin{verbatim}
$ N = $((2**24))
Benchmark
$ backend = "Metal"
$ ./runner_scripts/gpu/run_ode_mult_device.sh\
$N $backend
\end{verbatim}

\subsubsection{Benchmarking C++ (MPGOS) ODE solvers}

Benchmarking MPGOS ODE solvers requires the CUDA C++ compiler to be installed correctly.
The recommended CUDA Toolkit version is $\geq$11. The installation can be checked through

\begin{verbatim}
$ nvcc
If the installation exists, it will return 
something like this:
nvcc fatal   : No input files specified; 
use option --help for more information
\end{verbatim}

If \texttt{nvcc} is not found, the user needs to install CUDA Toolkit. The NVIDIA website lists out the resource \href{https://developer.nvidia.com/cuda-downloads}{\nolinkurl{https://developer.nvidia.com/cuda-downloads}} for installation.

The MPGOS scripts are in the \texttt{GPU\_ODE\_MPGOS} folder. The main executed code is the file \path{GPU\_ODE\_MPGOS/Lorenz.cu}. However, the MPGOS programs can be run with the same bash script by changing the arguments as
\begin{verbatim}
$ bash ./run_benchmark.sh -l cpp -d gpu -m ode
\end{verbatim}
which will generate the data files in \texttt{data/cpp} folder.

\subsubsection{Benchmarking JAX (Diffrax) ODE solvers}

Benchmarking JAX-based ODE solvers require installing Python 3.9 and \texttt{conda}. First, install all the Python packages for benchmarking:
\begin{verbatim}
$ conda env create -f environment.yml
$ conda activate venv_jax
\end{verbatim}
which should install the correct version of JAX with CUDA enabled and the Diffrax library. The GitHub \href{https://github.com/google/jax\#installation}{\nolinkurl{https://github.com/google/jax\#installation}} is a guide to follow if the installation fails.

For our purposes, the solvers can be benchmarked by
\begin{verbatim}
$ bash ./run_benchmark.sh -l jax -d gpu -m ode
\end{verbatim}

\subsubsection{Benchmarking PyTorch (torchdiffeq) ODE solvers}

Benchmarking PyTorch-based ODE solvers is a similar process as for JAX:
\begin{verbatim}
$ conda env create -f environment.yml
$ conda activate venv_torch
\end{verbatim}

\texttt{torchdiffeq} does not fully support vectorized maps with ODE solvers. To circumvent this, we extended the functionality by rewriting some library parts. To download the revision:
\begin{verbatim}
(venv_torch)$ pip uninstall torchdiffeq
(venv_torch)$ pip uninstall torchdiffeq
(venv_torch)$ pip install git+https://github.com/utkarsh530/torchdiffeq.git@u/vmap
\end{verbatim}
Then run the benchmarks by
\begin{verbatim}
$ bash ./run_benchmark.sh -l pytorch -d gpu -m ode
\end{verbatim}

\subsection{Comparing GPU acceleration of ODEs with CPUs}

The benchmark suite can also be used to test the GPU acceleration of ODE solvers in comparison with CPUs. The process for generating simulation times for GPUs can be done by following Section \ref{subsubsection: bench_julia}. CPU simulation times for ODEs can be generated by the bash script:
\begin{verbatim}
$ bash ./run_benchmark.sh -l julia -d cpu -m ode
\end{verbatim}

The simulation times will be generated in \texttt{data/CPU}. Each of the workflows takes about 20 minutes to finish.

\subsection{Benchmarking GPU acceleration of SDEs with CPUs}

The SDE solvers in Julia are benchmarked by comparison to the CPU-accelerated simulation. 
Generate simulation times for GPU by using
\begin{verbatim}
$ bash ./run_benchmark.sh -l julia -d gpu -m sde
\end{verbatim}

Generate the simulation times for CPU-accelerated codes through
\begin{verbatim}
$ bash ./run_benchmark.sh -p julia -d cpu -m sde
\end{verbatim}

The results will be generated in \texttt{data/SDE} and \texttt{data/CPU/SDE}, taking about 10 minutes to complete.

\subsection{Composability with MPI}

Julia supports Message Passing Interface (MPI) to allow Single Program Multiple Data (SPMD)-type parallel programming. The composability of the GPU ODE solvers enables seamless integration with MPI, enabling scaling the ODE solvers to clusters on multiple nodes:
\begin{verbatim}
$ julia --project=./GPU_ODE_Julia
julia> using Pkg
# install MPI.jl
julia> Pkg.add("MPI")
\end{verbatim}

An example script solving the Lorenz problem for approximately 1 billion parameters is available in the \texttt{MPI} folder. A SLURM-based script is 
\begin{verbatim}
#!/bin/bash
# Slurm Sbatch Options
# Reqeust no. of GPUs/node
#SBATCH --gres=gpu:volta:1
# 1 process per node 
#SBATCH -n 5 -N 5
#SBATCH --output="./mpi_scatter_test.log-%j"
# Loading the required module

# MPI.jl requires memory pool disabled
export JULIA_CUDA_MEMORY_POOL=none
export JULIA_MPI_BINARY=system
# Use local CUDA toolkit installation
export JULIA_CUDA_USE_BINARYBUILDER=false

source $HOME/.bashrc
module load cuda mpi

srun hostname > hostfile
time mpiexec julia --project=./GPU_ODE_Julia ./MPI/gpu_ode_mpi.jl
\end{verbatim}

\subsection{Plotting Results}

The plotting scripts to visualize the simulation times are located in the \texttt{runner\_scripts/plot} folder. These scripts replicate the benchmark figures in the paper. The benchmark suite contains the simulation data generated by authors, which can be used to verify the plots. Various benchmarks can be plotted, which are described in the different sections. The plotting scripts are based on Julia. As a preliminary step:

\begin{verbatim}
$ cd GPUODEBenchmarks
$ julia project=.
julia> using Pkg
julia> Pkg.instantiate()
julia> Pkg.precompile()
\end{verbatim}

The plot comparison between Julia, C++, JAX, and PyTorch mentioned in the paper can be generated by using the command:
\begin{verbatim}
$ julia --project=. ./runner_scripts/plot/plot_ode_comp.jl
\end{verbatim}

The plot will be saved in \texttt{plots} folder. 

Similarly, the other plots in the paper can be generated by running the different scripts in the folder \texttt{runner\_scripts/plot}.

\begin{verbatim}
plot performance of GPU ODE solvers 
with multiple backends
$ julia --project=. ./runner_scripts/plot/plot_mult_gpu.jl 
plot GPU ODE solvers comparsion with CPUs
$ julia --project=. ./runner_scripts/plot/plot_ode_comp.jl 
plot GPU SDE solvers comparsion with CPUs
$ julia --project=. ./runner_scripts/plot/plot_sde_comp.jl 
plot CRN Network sim comparsion with CPUs
$ julia --project=. ./runner_scripts/plot/plot_sde_crn.jl 
\end{verbatim}

To plot data generated by running the scripts, specify the location of the \texttt{data} as the argument to the mentioned command:
\begin{verbatim}
$ julia --project=. ./runner_scripts/plot/plot_mult_gpu.jl /path/to/data/
\end{verbatim}
\end{document}